\documentclass[lettersize,journal]{IEEEtran}
\usepackage{booktabs}
\usepackage{multirow}
\usepackage{cite}
\usepackage{url}
\usepackage{ragged2e}
\usepackage{footnote}
\makesavenoteenv{tabular}
\makesavenoteenv{table}
\usepackage{epsfig}
\usepackage{graphicx}
\usepackage{amsmath,amssymb} % define this before the line numbering.
\usepackage{algorithm}
\usepackage{algorithmicx}
\usepackage{algpseudocode}
\usepackage{graphics}
\usepackage{threeparttable}
\usepackage{color}
\usepackage[normalem]{ulem}
\usepackage{multirow}
\usepackage{float}
\usepackage{amsfonts}
\usepackage{bm}
\usepackage{array}
\usepackage{pifont}
\usepackage{diagbox}
\usepackage{rotating}
\usepackage{booktabs}
\usepackage{overpic}
\usepackage{textcomp}
\usepackage{contour}
\let\labelindent\relax
\usepackage{enumitem}
\usepackage{colortbl}
\usepackage{csquotes}
\usepackage{xr}
\usepackage[american]{babel}
\usepackage{microtype}
\usepackage{bbding}
\usepackage{amsfonts,amssymb}
\usepackage[pagebackref=false,breaklinks=true,colorlinks, bookmarks=false]{hyperref}
\usepackage[table,xcdraw]{xcolor}
\usepackage{colortbl}
\usepackage{cleveref}
\usepackage{arydshln}
\crefformat{section}{\S#2#1#3} 
\crefformat{subsection}{\S#2#1#3}
\crefformat{subsubsection}{\S#2#1#3}
\usepackage{silence}
\usepackage{booktabs}
\usepackage{makecell}
\def\ie{\emph{i.e.}}
\def\eg{\emph{e.g.}}

\def\etal{{\em et al.~}}

\newcommand{\figref}[1]{Fig.~\ref{#1}}
\newcommand{\tabref}[1]{Table~\ref{#1}}

\usepackage{tabularx}
\def\ie{\emph{i.e.}}
\def\eg{\emph{e.g.}}

\def\etal{{\em et al.~}}

\usepackage{colortbl}
\usepackage{cleveref}
\crefformat{section}{\S#2#1#3} 
\crefformat{subsection}{\S#2#1#3}
\crefformat{subsubsection}{\S#2#1#3}
\def\BibTeX{{\rm B\kern-.05em{\sc i\kern-.025em b}\kern-.08em
    T\kern-.1667em\lower.7ex\hbox{E}\kern-.125emX}}
\usepackage{balance}
\begin{document}
\title{
Patch Triplet Similarity Purification for Guided Real-World Low-Dose CT Image Denoising}
\author{Junhao Long, Fengwei Yang, Juncheng Yan, Baoping Zhang, Chao Jin, Jian Yang, Changliang Zou, Jun Xu 
\thanks{This research is supported by The National Natural Science Foundation of China (No. 12226007, 62176068, and 62171309), the Open Research Fund from the Guangdong Provincial Key Laboratory of Big Data Computing, The Chinese University of Hong Kong, Shenzhen, under Grant No. B10120210117-OF03, Youth Project of Tianjin Municipal Applied Basic Research Project (23JCQNJC01630), and the Fundamental Research Funds for the Central Universities.
Corresponding author: Jun Xu (csjunxu@nankai.edu.cn) and Juncheng Yan (yanjuncheng@mail.nankai.edu.cn)
}
\thanks{J.-H. Long, F.-W. Yang, J.-C. Yan, C.-L. Zou, and J. Xu are with School of Statistics and Data Science, Nankai University, Tianjin, 300071, China.}
\thanks{B.-P. Zhang, J. Chao and J. Yang are with the Department of Radiology, The First Affiliated Hospital of Xi'an Jiaotong University, Xi'an, China; Shanxi Engineering Research Center of Computational Imaging and Medical Intelligence, Xi’an, China and Xi’an Key Laboratory of Medical Computational Imaging, Xi’an, China.
}
%\thanks{J. X. is with School of Statistics and Data Science, Nankai University, Tianjin, China and Guangdong Provincial Key Laboratory of Big Data Computing, The Chinese University of Hong Kong, Shenzhen, China.}
}
\maketitle

\begin{abstract}
Image denoising of low-dose computed tomography (LDCT) is an important problem for clinical diagnosis with reduced radiation exposure. Previous methods are mostly trained with pairs of synthetic or misaligned LDCT and normal-dose CT (NDCT) images. However, trained with synthetic noise or misaligned LDCT/NDCT image pairs, the denoising networks would suffer from blurry structure or motion artifacts. Since non-contrast CT (NCCT) images share the content characteristics to the corresponding NDCT images in a three-phase scan, they can potentially provide useful information for real-world LDCT image denoising. To exploit this aspect, in this paper, we propose to incorporate clean NCCT images as useful guidance for the learning of real-world LDCT image denoising networks. To alleviate the issue of spatial misalignment in training data, we design a new Patch Triplet Similarity Purification (PTSP) strategy to select highly similar patch (instead of image) triplets of LDCT, NDCT, and NCCT images for network training. Furthermore, we modify two image denoising transformers of SwinIR and HAT to accommodate the NCCT image guidance, by replacing vanilla self-attention with cross-attention. On our collected clinical dataset, the modified transformers trained with the data selected by our PTSP strategy show better performance than 15 comparison methods on real-world LDCT image denoising. Ablation studies validate the effectiveness of our NCCT image guidance and PTSP strategy. We will publicly release our data and code.
\end{abstract}
%yfw：1.我看一些文章对于辐射的表述是reduced radiation exposure
%2.两句都在说synthetic or misaligned LDCT/NDCT image pairs，是不是可以合并一下第二句，不重复说主语了
%3.感觉To alleviate the issue of能不能用To address来替代呢
%4.感觉这句子有点长，要不要把to select...改成从句
%5，感觉加一些这种词突出结构感
%6.句子有点长，不知道这样精简一下可不可以

\section{Introduction}
\label{sec:introduction}

\begin{figure}[!t]
\centerline{\includegraphics[width=\columnwidth]{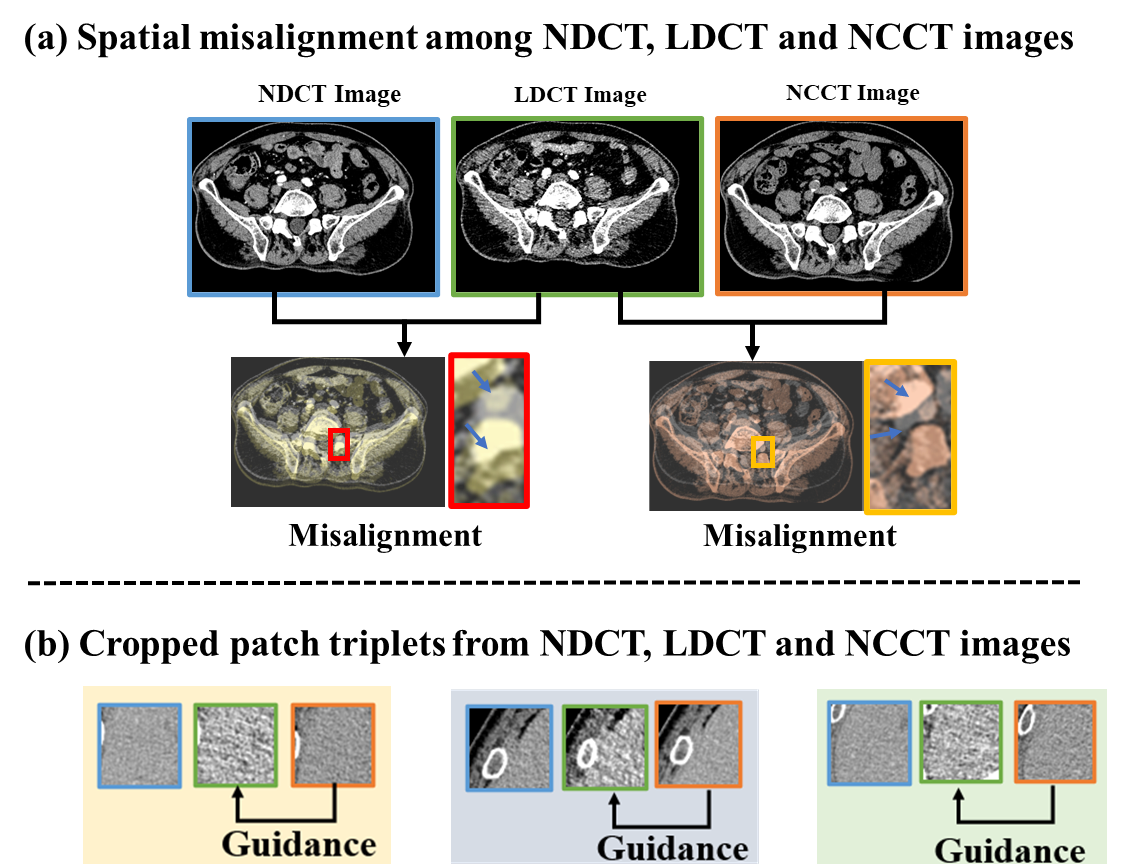}}
\caption{\textbf{Motivation of our NCCT image guidance and Patch Triplet Similarity Purification (PTSP) strategy}.
(a) The NCCT image enjoys structural similarity with the corresponding NDCT images from three-phase scanning. However, when overlapping LDCT images with the corresponding NDCT and NCCT images, there exists clear spatial misalignment.
(b) Utilizing patch-level guidance instead of image-level one for network training.
}
\label{fig:mismatch}
\end{figure}

\IEEEPARstart{T}{he} technology of computerized tomography (CT) scanning is widely used for clinical diagnosis~\cite{hricak2011managing,raman2013ct,karatas2014three}. For example, the annual frequency of CT examinations was 239.8 per 1000 inhabitants~\cite{ctfrequency}. However, the exposure to radiation by X-rays in CT scanning brings potential health risks to the human body~\cite{cancer2024}. To alleviate this problem, it is essential to reduce the dose usage of CT scanning in clinical diagnosis. But low-dose CT (LDCT) images usually suffer from unclear details due to noise and artifacts, making it difficult for physicians to accurately diagnose lesions. Thus, LDCT image denoising becomes a meaningful research topic.
%ljh:直接叙述低剂量CT图像造成图片质量下降，影响医生判断，然后引出降噪的重要性

During the past decade, many LDCT image denoising methods have been developed based on deep convolutional neural networks (CNN)~\cite{chen2017low,CZZ,kang2018deep} or transformers~\cite{zm,li2022transformer,dosovitskiy2020image}.
These works mainly tackle synthetic LDCT images generated by adding Poisson noise to the sinogram data of the corresponding normal-dose CT (NDCT) images~\cite{chen2017low}.
But the denoising networks trained with synthetic noise which is really different from the noise of the real-world LDCT images usually perform poorly in removing realistic noise~\cite{xu2020noisy,mjh,zdq}.
To address real-world noise in clinical scenarios, several recent methods~\cite{noise2noise,krull2019noise2void,batson2019noise2self,xu2020noisy} perform LDCT image denoising under self-supervised learning frameworks~\cite{SSL,hou2018image}. However, without supervision of the NDCT images, it is still difficult to remove complex noise from the real-world LDCT images~\cite{chen2017low,xu2017,wang2023ctformer}.
%  (\figref{fig:methods} (c) and (g))
%有的用合成数据，有的用自监督，都存在问题，接下来就介绍使用真实数据的有监督的也存在问题，引入guidance
%临床实践->real-world  complex noise说明自监督难以胜任从而引出GAN
For real-world LDCT image denoising, many methods~\cite{yang2018low,li2023low,bera2021noise} employ generative adversarial networks (GAN)~\cite{gan,zhu2017unpaired} to learn direct mappings between real-world LDCT images and their spatially similar but misaligned NDCT counterparts (\figref{fig:mismatch}). However, using such ``pairs'' of LDCT and NDCT images for network training would result in structural distortions in the restored images. 
%(\figref{fig:methods}).
% In addition, obtaining such ``paired'' LDCT and NDCT images often requires a lot of time and manpower.

%yfw:Alternatively引出第二种选择或可能的建议，或者类似的词，可能会让结构更清晰
%把goodfellow2020generative改成gan了

% 先说motivation再说创新点，再说具体怎么做的。
% 第一个点是为了解决真实LDCT和NDCT空间不匹配的问题，设计了PTSP的策略去筛选训练数据；考虑到平扫数据能带来额外的结构信息，因此引入cross attention到已有的流行图像去噪transformer里做guided LDCT image denoising；构建了一个真实数据集，包括LDCT，NDCT和NCCT，可以用来做什么
%根据8月21号讨论修改逻辑，ljh修改如下：先说平扫能够提供额外信息，但是存在空间不匹配的问题，所以要引入PTSP策略

% 1.for guidance, we need NCCT
In clinical practice, the non-contrast CT (NCCT) images are usually scanned to provide vascularization characteristics and enhanced lesion patterns of the LDCT images. As shown in \figref{fig:mismatch}, for an LDCT image, its NCCT image has similar structure to its NDCT image. This indicates that NCCT images can provide helpful guidance on LDCT image denoising which is ignored by past researches.
%2. describe overall objective and method
To exploit this aspect, in this paper, we propose to modify popular image denoising transformers~\cite{SwinIR,chen2023hat} to accommodate with ``triplets'' of LDCT, NDCT, and NCCT images.
%3. for NCCT guided LDCT image denoising, we need collect training data of patch triplets 
However, it is necessary to minimize the side effects of spatial misalignment among the LDCT, NDCT, and NCCT images (\figref{fig:mismatch}) on training real-world LDCT image denoising networks.
% But there is noise between LDCT images and NDCT images as well as NCCT images. This noise can cause "brightness" differences among the images, as shown in \figref{motivation}, making it difficult to assess structural similarity.
For this goal, we propose a Patch Triplet Similarity Purification (PTSP) strategy to select highly similar NDCT and NCCT image patches as the ``target'' and guidance reference, respectively, of each LDCT image patch at the same locations of corresponding images.
Our PTSP strategy is built upon pixel value discretization~\cite{psp2024} to be robust on noise degradation in LDCT images.

%4. with our data, cross-attention is imlemented on two methods 
With highly similar training data selected by our PTSP strategy, we employ cross-attention~\cite{Chen_2021_ICCV} to incorporate useful information from clean NCCT images into image denoising transformers for guided LDCT image denoising.
%5. experiments show our advantages 
%6. ablation study validate the effectivess of our contributions.
Experiments on our synthetic and clinical datasets demonstrate that, with our NCCT guidance and PTSP strategy, the modified SwinIR~\cite{SwinIR} and HAT~\cite{chen2023hat} obtain better LDCT image denoising performance than 15 comparison methods. Ablation studies validate the effectiveness of our NCCT image guidance and PTSP strategy in selecting high-quality training triplets of LDCT, NDCT, and NCCT image patches for LDCT image denoising.

In summary, our main contributions are three-fold:
\begin{itemize}
    \item To exploit extra information from clean NCCT images, \textbf{we propose to incorporate NCCT image as useful guidance for real-world LDCT image denoising}. This is implemented by replacing vanilla self-attention with cross-attention in image denoising transformers.
    
    \item To address the spatial misalignment between real-world LDCT images and NDCT/NCCT images, \textbf{we propose a Patch Triplet Similarity Purification (PTSP) strategy to select highly similar triplets of LDCT, NDCT, and NCCT image patches with negligible misalignment} as high-quality training data for LDCT image denoising.

     \item \textbf{Incorporated by the guidance from NCCT images, two transformers~\cite{SwinIR,chen2023hat} modified to be trained using our PTSP strategy outperform fifteen LDCT image denoising methods} on our collected clinical dataset. 
     % Ablation studies also investigate the hyper-parameters of our Guided Transformer Network and PTSP strategy.
\end{itemize}

The remaining parts of this paper are organized as follows.
We present the related work in \cref{sec:relatedwork}.
In \cref{sec:method}, we propose our PTSP strategy and NCCT image guidance.
In \cref{sec:dataset}, we provide the introduction of our synthetic dataset and clinical dataset.
Experiments in \cref{sec:experiment} demonstrated that our PTSP strategy and NCCT image guidance boosts two LDCT denoising networks both quantitatively and qualitatively.
The conclusion is summarized in \cref{sec:conclusion}.

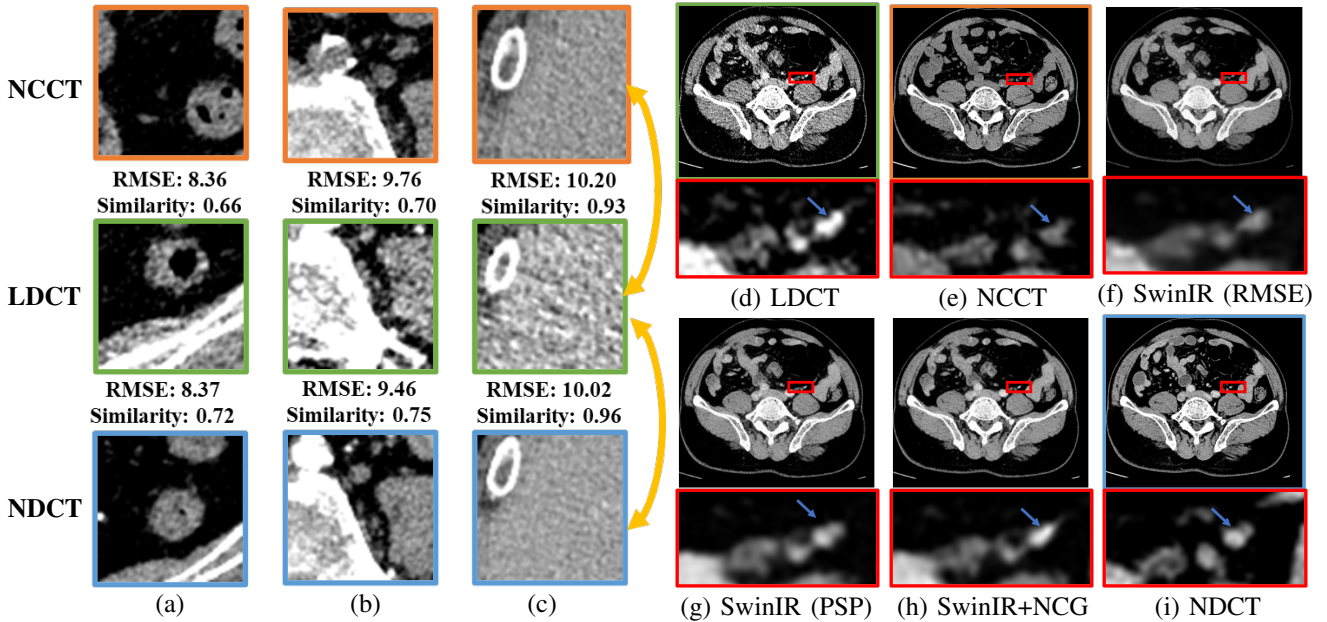
\begin{figure*}[t]
  \centering
  \begin{minipage}[b]{0.433\textwidth}
    \begin{overpic}[width=\textwidth]{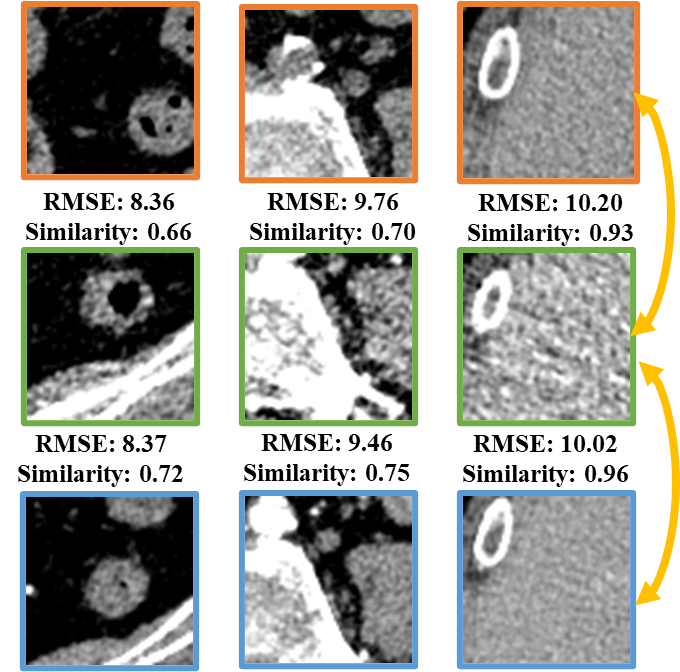}
      \put(12,-4){ (a)}
      \put(45,-4){ (b)}
      \put(75,-4){ (c)}
      \put(-13,48){ \textbf{LDCT}}
      \put(-13,12){ \textbf{NDCT}}
      \put(-13,83){ \textbf{NCCT}}
    \end{overpic}
   \end{minipage}
  \begin{minipage}[b]{0.15\textwidth}
  
    \begin{overpic}[width=\textwidth]{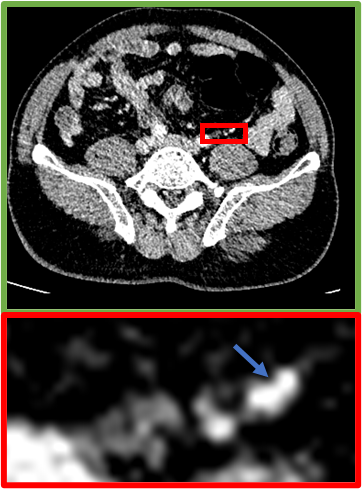}
      \put(20,-10){(d) LDCT}
    \end{overpic}
    \vskip 15pt % 增加间距
    \begin{overpic}[width=\textwidth]{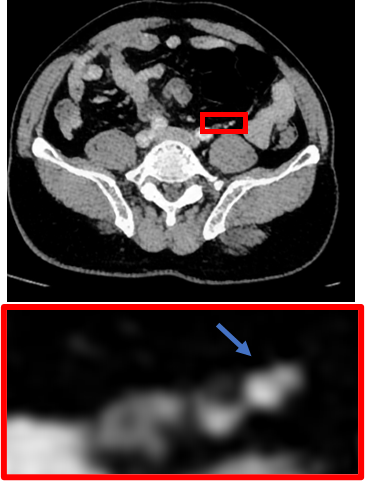}
      \put(2,-10){(g) SwinIR (PSP)}
    \end{overpic}
  \end{minipage}
  \begin{minipage}[b]{0.15\textwidth}
    \begin{overpic}[width=\textwidth]{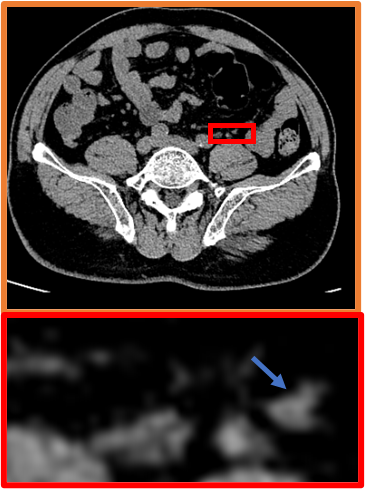}
      \put(18,-10){(e) NCCT}
    \end{overpic}
    \vskip 15pt % 增加间距
    \begin{overpic}[width=\textwidth]{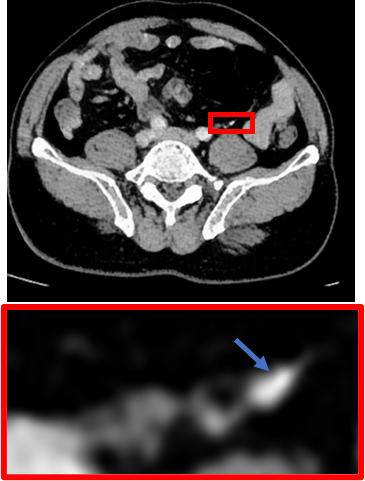}
      \put(3,-10){(h) SwinIR+NCG}
    \end{overpic}
  \end{minipage}
  \begin{minipage}[b]{0.15\textwidth}
    \begin{overpic}[width=\textwidth]{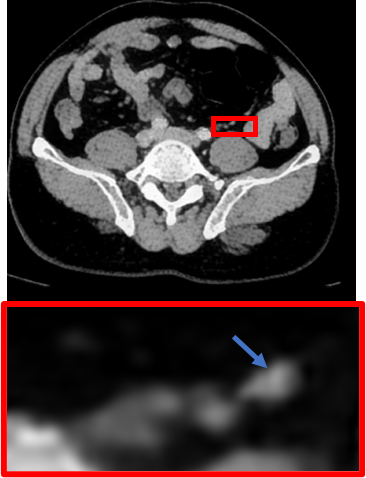}
      \put(-2,-10){(f) SwinIR (RMSE)}
    \end{overpic}
    \vskip 15pt % 增加间距
    \begin{overpic}[width=\textwidth]{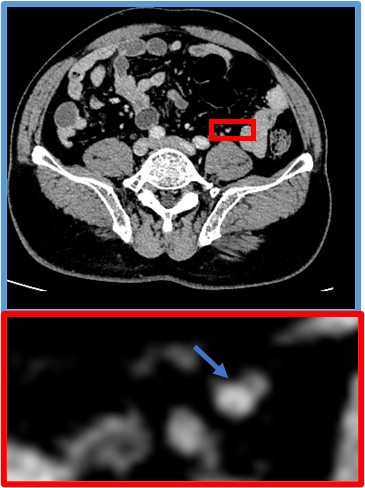}
      \put(19,-10){(i) NDCT}
    \end{overpic}
  \end{minipage}
\vspace{1mm}  \caption{\textbf{Training data screening strategy based on RMSE \textsl{v.s.} our Patch Triplet Similarity Purification (PTSP) strategy}. Mask Similarity is abbreviated as ``Similarity''. Subfigure (a) shows more significant structure differences among LDCT, NDCT, and NCCT image patches than (b) and (c). However, the RMSE metric gives the opposite conclusion. There are obvious ``brightness'' differences among the three image patches in (c). The average pixel value of the LDCT patch is 169.23, which is 25.26 higher than the average pixel value of the NCCT patch and 10.64 higher than that of the NDCT patch. (d) An LDCT image. (e) The corresponding NCCT image. (f) The denoised image of SwinIR~\cite{SwinIR} trained with the data screened by the RMSE metric (g) The denoised image of SwinIR by introducing PSP strategy~\cite{psp2024}.
    (h) The denoised image of SwinIR by introducing NCCT image guidance and our PTSP strategy.
    (i) The corresponding NDCT image.
    In general, the proposed NCCT image guidance and PTSP strategy for training data selection well recover the structure of the denoised image on real-world LDCT image denoising.}
    \vspace{-4mm}
  \label{fig:motivation}
\end{figure*}

\section{Related Work}
\label{sec:relatedwork}
In this section, we introduce the work closely related to ours, including the LDCT image denoising methods in \cref{subsec:LDCT reconstruction}, self-supervised image denoising method in \cref{subsec:unpaired}, and guided image denoising in \cref{subsec:guided}.

\subsection{Low-Dose CT Image Denoising}
\label{subsec:LDCT reconstruction}
Low-dose CT (LDCT) image denoising is initially tackled with first convolutional neural networks (CNNs)~\cite{CZZ}. The RED-CNN network~\cite{chen2017low} was developed with an encoder-decoder architecture for favorable performance. Kang \etal~\cite{kang2017deep,kang2018deep} proposed to learn wavelet transforms with CNNs for LDCT image denoising.
%
%yfw: firstly好像是首先/第一的意思，这里是不是想表达“最初”，可能initially或者at first比较合适？
Compared to CNNs~\cite{TMI1}, transformers~\cite{transformer} are good at capturing global information and long-range feature interactions, which have been applied to LDCT image denoising for better performance~\cite{wang2023ctformer}. Vision Transformer (ViT)~\cite{dosovitskiy2020image} has also been utilized in~\cite{wang2021ted,wang2023ctformer} to enlarge the effective receptive fields of window-based transformers for better denoising performance. Li \etal~\cite{li2022transformer} devised a dual-branch transformer to recover the edges and textures of LDCT images well.
%CNNs和Transformer用的都是合成的低剂量CT，而下面的GAN用的是real-world的
However, the above methods are trained with synthetic LDCT images and could hardly be applied to real-world LDCT images in clinical practice.
%by adding Poisson noise into the sinograms of the corresponding normal-dose CT (NDCT) images. 

% routine-dose
For clinical purposes, researchers proposed to learn mapping from real-world LDCT images to high-quality NDCT ones using the GAN architectures~\cite{gan}. Wolterink \etal~\cite{wolterink2017generative} trained a CNN jointly with an adversarial CNN to recover the NDCT images from LDCT images. Yi and Babyn~\cite{yi2018sharpness} trained an adversarial network together with a sharpness detection network to mitigate the blurring effects in LDCT image denoising. Later, many LDCT image denoising methods are built upon CycleGAN~\cite{kang2019cycle,tang2019unpaired}, conditional GAN~\cite{hong2020end}, or WGAN~\cite{yang2018low,hu2019artifact}. However, the structural misalignment between real-world LDCT images and NDCT images makes it difficult to guarantee the fidelity of denoised images~\cite{kulathilake2023review}.

%两点不同：(1)放弃使用合成数据(2)用PTSP进行筛选
In this paper, we also use real-world LDCT images to train the denoising networks, thereby well serving clinical practice. To alleviate the problem of image structure misalignment between LDCT and NDCT images, we propose a Patch Triplet Similarity Purification (PTSP) strategy to select highly-similar patch triplets for training LDCT image denoising networks.

\subsection{Self-Supervised Image Denoising}
\label{subsec:unpaired}
Self-supervised image denoising methods~\cite{noise2noise,xu2020noisy} learn from the noisy images themselves to remove the noise without using clean images.
By assuming that noise is zero-mean and independently and identically distributed (i.i.d.), Noise2Noise (N2N)~\cite{noise2noise} effectively trains image denoising using pairs of noisy images with the same contents but different noise. Noise2Void (N2V)~\cite{krull2019noise2void} learns to predict the true value of each noisy pixel from its neighboring pixels, and hence called ``blind-spot'' method.
Unlike N2V, Noise2Self (N2S)~\cite{batson2019noise2self} additionally performs masking operations for each pixel to enhance the denoising robustness.
Noise-As-Clean (NAC)~\cite{xu2020noisy} learns to remove image noise with a pair of noisy image and noisier image, which is produced by adding synthetic noise to the noisy image.
However, the above-mentioned denoising methods mainly learn to remove the zero-mean and i.i.d. noise, which may not hold true for real-world LDCT images.

Self-supervised learning has also been applied to LDCT image denoising by only using LDCT images.
%原来的说法是论文的摘要里面提炼的，说的不够详细，感觉直接使用原文的摘要里面的说法更清楚一点：
Noise2Inverse~\cite{hendriksen2020noise2inverse} performs image reconstruction~\cite{TMI2} by learning a CNN without additional clean or noisy data.
%这篇论文的similarity是根据RMSE指标和选定的阈值来形成掩码从而使得只有像素值接近的区域才能够纳入优化的范围
Noise2Sim~\cite{niu2022noise} is a self-supervised deep denoising approach that achieves noise reduction by using similar images.
However, because of the lack of supervision from high-quality NDCT images, it is challenging for these self-supervised denoisers to remove complex noise well in real-world LDCT images.

In this paper,
%unlike self-supervised methods that train denoising networks only using LDCT images, 
we propose to train denoising networks with pairs of highly similar LDCT and NDCT image patches selected by our similarity purification strategy.
% supervised learning scheme

\subsection{Guided Image Denoising}
\label{subsec:guided}
%下面的这些方法中只有MLEFGN和MLEFGN是CT降噪的，因此将MLEFGN放到后面，其他的放到前面来

Many image denoising methods utilize useful spatial or edge information from external clean images for guided image denoising. He \etal~\cite{He} proposed guided image filtering (GIF) to use the guidance image to identify noise and edges for better noise reduction. Based on GIF, the method of~\cite{WGIF} incorporates an edge-aware weighting strategy for edge-preserving image filtering. Xu \etal~\cite{NSS} exploited the external information from clean images to guide the internal learning of a noisy test image for real-world image denoising. Zhang \etal~\cite{zhang2022guided} utilized the mean image of all spectral bands as useful guidance to adaptively aggregate spatial information.

The insights of external guidance also boost LDCT image denoising. For example, edge-guided filtering~\cite{edge-guided} and GDAFormer~\cite{Jiang2024GDAFormerGD} use edge feature to guide the learning of LDCT image denoising.
%说明缺陷，引出NCCT image guidance
%这些缺陷有点牵强，并且我们也using external data了 However, using edge features as a guide provides too little information, while using external data as a guide makes it difficult to ensure the consistency of the distribution between external and internal data.
In this paper, we use Non-Contrast CT (NCCT) image to guide the LDCT image denoising.

\section{Proposed Method}
\label{sec:method}

In this section, we propose a NCCT image guidance for LDCT image denoising in \cref{subsec:NCCT}. Then we introduce our Patch Triplet Similarity Purification (PTSP) strategy in \cref{subsec:PTSP}, to select highly similar training data. With our PTSP strategy, we integrate the guidance of NCCT images into two denoising Transformers for LDCT image denoising in~\cref{subsec:training}.

\begin{figure*}[!t]
\centerline{\includegraphics[width=\textwidth]{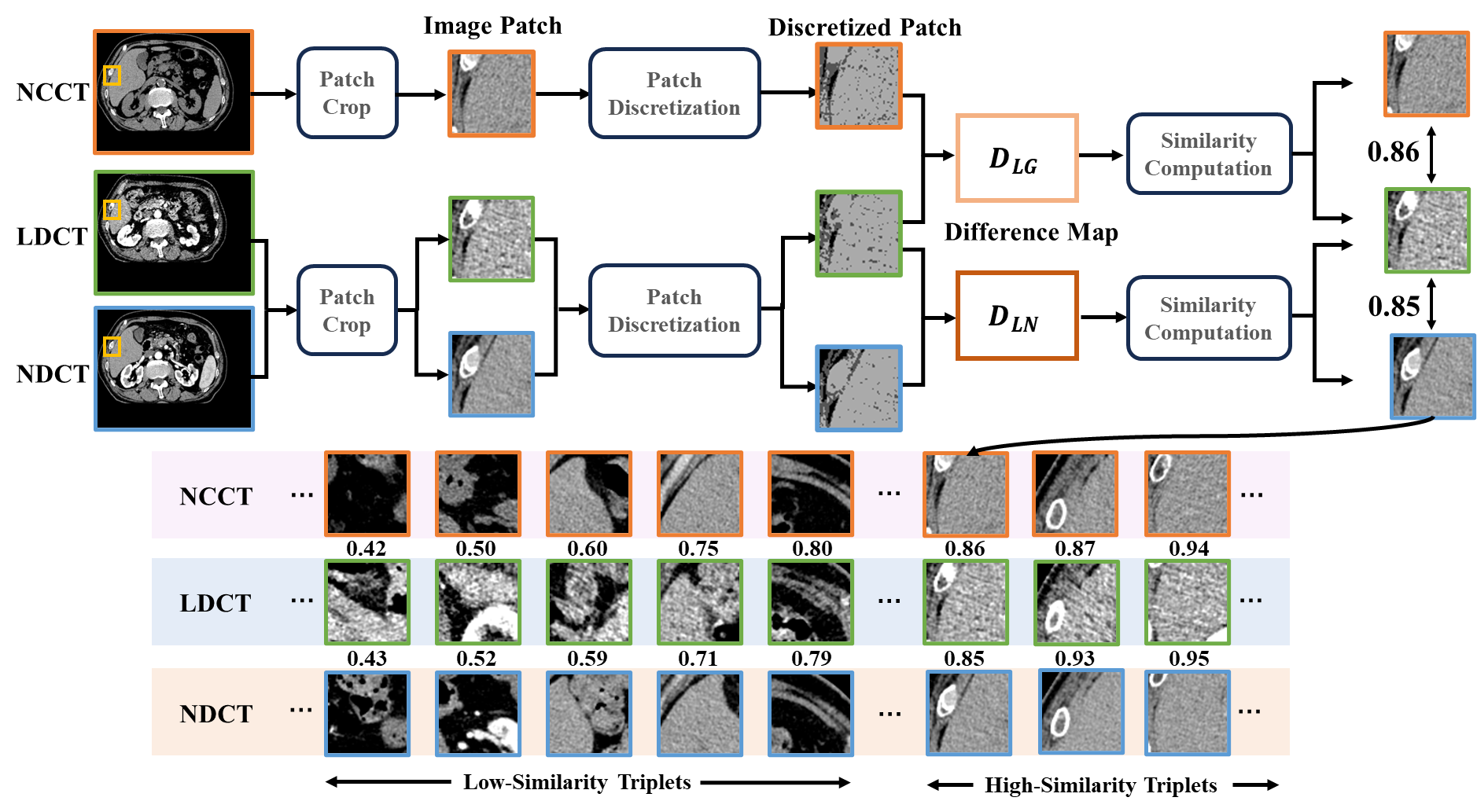}}
\vspace{-3mm}
\caption{\textbf{The proposed Patch Triplet Similarity Purification (PTSP) Strategy}. It includes three main steps: 
1) compute the discretized image patches $M_G$, $M_L$, and $M_N$ according to the set pixel interval; 
2) obtain the difference maps $D_{LN}$ (or $D_{LG}$) by subtracting the discretized LDCT image patch from the discretized NDCT (or NCCT) image patch; 
3) compute the corresponding mask similarity based on difference maps. When the mask similarity reaches a preset threshold $s$ (\eg, $s=0.85$), we include it in the training set of our clinical dataset.
}
\label{fig:ptsp_strategy}
\vspace{-3mm}
\end{figure*}

\subsection{NCCT-Guided LDCT Image Denoising}
\label{subsec:NCCT}
For LDCT image denoising, the self-attention-based transformers~\cite{Jiang2024GDAFormerGD} achieve promising performance when trained with LDCT and NDCT images. However, this may inaccurately estimate the tissue structure of LDCT images with noise degradation.
What's more, these methods ignored the useful information from the NCCT images, which are widely used for preliminary diagnosis of diseases in clinical practice~\cite{van2018skeletal}. As shown in \figref{fig:mismatch}, the NCCT image has similar structure and texture to the corresponding NDCT image, which provides useful information for LDCT image denoising. Inspired by this observation, we propose to utilize the clean NCCT images as complementary guidance to remove the noise from LDCT images. This guidance can be implemented in a cross-attention mechanism~\cite{Chen_2021_ICCV} for transformer based denoisers~\cite{SwinIR,chen2023hat}, where the LDCT image provides the query matrix $\mathbf{Q}$ and value matrix $\mathbf{V}$ while the NCCT image provides the key matrix $\mathbf{K}$.
This allows cross-attention to establish the associations between the noisy LDCT image and the corresponding clean NCCT image, enhancing the capability of the denoising transformers on structure preserving and texture recovery.

Direct using triplets of LDCT, NDCT, and NCCT images for network training does not bring promising performance on LDCT image denoising. The key problem is that the images from the three-phase scanning suffer from clear spatial misalignment (\figref{fig:mismatch}). To alleviate this issue, we propose to select highly similar patch triplets (instead of image triplets) for network training, as will be introduced as follows.

\subsection{Proposed Patch Triplet Similarity Purification Strategy}
\label{subsec:PTSP}

To address spatial misalignment between LDCT and NDCT images, a natural idea is to select pairs of highly similar LDCT and NDCT images for network training. Similarity can be measured using the RMSE metric~\cite{buades2005non,dabov2007image,xu2018trilateral,nonlocal}. That is, for each reference LDCT patch, these methods search for the most similar patch to it from the NDCT image as the training ``target''. However, the RMSE metric is error-prone in measuring the similarity between LDCT and NDCT image patches, since the distance is largely influenced by the noise in LDCT images and the misalignment between LDCT and NDCT images (\figref{fig:motivation}).
%Besides, it is computationally expensive to select the most similar image patch from the NDCT images for all LDCT patches.
For example, the patches in \figref{fig:motivation} (c) are more similar to each other from the perspective of visual effects than those in \figref{fig:motivation} (a) and (b). However, the RMSE distances could not reflect this trend. Training transformers using pairs of similar LDCT and NDCT patches selected by minimal RMSE distance would result in vague structure or visual distortions in the denoised images like \figref{fig:motivation} (f). In contrast, transformers trained with the introduction of NCCT image guidance and our PTSP strategy effectively alleviate this issue, as shown in~\figref{fig:motivation} (h).

To provide high-quality training data for guided LDCT image denoising, in this work, we propose a Patch Triplet Similarity Purification (PTSP) strategy to select highly similar patch triplets from clinical LDCT, NDCT, and NCCT images with consistent tissue structures, which is shown in \cref{fig:ptsp_strategy}.
%The selected patch triplets are used for the training of LDCT image denoising networks.
For each patch triplet, our PTSP strategy contains three main steps: 1) discretizing image patches according to the pixel intervals; 2) obtaining difference maps by subtracting the discretized LDCT patch from the discretized NDCT patch or NCCT patch; 3) computing the corresponding mask similarity based on the difference maps and obtaining the training patch triplets with high similarity.

\noindent
\textbf{Patch discretization}.
%\xj{the goal of this process; given an LDCT image, how to discretize}
This step aims to describe the tissue content of each patch for similarity computation. The description of tissue contents is implemented by discretizing the pixel values of LDCT image patch, NDCT image patch, and NCCT image patch into multiple segments. Specifically, the pixel values in each CT image patch are between 0 and 255, we divide them into multiple segments separated by a set of predefined points $\{T_i\}_{i=0}^{n}$ that satisfies: $0=T_0<T_1<...<T_{n}=256$, where $n$ is the number of segments. Denoting $x$ as the position in LDCT patch $\bm{p}_{L}$, NDCT patch $\bm{p}_{N}$, or NCCT patch $\bm{p}_{G}$ of size $p\times p$ ($p=64$ in our experiments), the discretized patch $\bm{M}_{b}$ ($b$ is ``$L$'', ``$N$'' or ``$G$'') is defined as:
\begin{equation}  
\bm{M}_b(x) =   
\begin{cases}  
0, &\text{if } T_{0} \leq \bm{p}_b(x) < T_1,\\
i-1, &\text{if } T_{i-1} \leq \bm{p}_b(x)<T_i, 1<i<n,\\
n-1, &\text{if } T_{n-1} \leq \bm{p}_b(x) < T_n. 
\end{cases}  
\label{discretization}  
\end{equation}  
The set of separation points $\{T_i\}_{i=0}^n$ and the segment number $n$ need to be predefined in advance. For example, when $n=3$, $\{T_i\}_{i=0}^n$ can be set as $\{0, 85, 170, 256\}$ based on a linear separation scheme. The separation points and the number of segments can be flexibly set based on the actual situation. 
%对于PSP策略最后去说
%For the network training without the guidance of NCCT image, the patch discretization only needs to be operated on pairs of LDCT and NDCT images.

\noindent
\textbf{Patch differentiation}.
The goal of this step is to measure the distance between pairs of LDCT and NDCT image patches as well as between pairs of LDCT and NCCT image patches. In this step, the difference map $\bm{D}_{LN}$ (or $\bm{D}_{LG}$) is obtained by subtracting the discretized NDCT patch $\bm{M}_{N}$ (or NCCT patch $\bm{M}_{G}$) from the discretized LDCT patch $\bm{M}_{L}$, as follows:
\begin{equation}  
\label{difference}  
\begin{aligned}  
\bm{D}_{LN}(x) &= \lvert \bm{M}_L(x) - \bm{M}_N(x) \rvert, \\  
\bm{D}_{LG}(x) &= \lvert \bm{M}_L(x) - \bm{M}_G(x) \rvert.  
\end{aligned}  
\end{equation}
The range of difference maps $\bm{D}_{LN}$ and $\bm{D}_{LG}$ is $0\sim n-1$.
Higher values at the position $x$ of difference map $\bm{D}_{LN}$ (or $\bm{D}_{LG}$) usually indicates larger difference between the corresponding pixel values of LDCT and NDCT (or NCCT) image patches at that pixel position $x$.
%If there is no guidance from the NCCT images, only the difference map between the LDCT and NDCT image patches needs to be calculated in the PSP strategy.
\begin{figure}[!t]
\centerline{\includegraphics[width=\columnwidth]{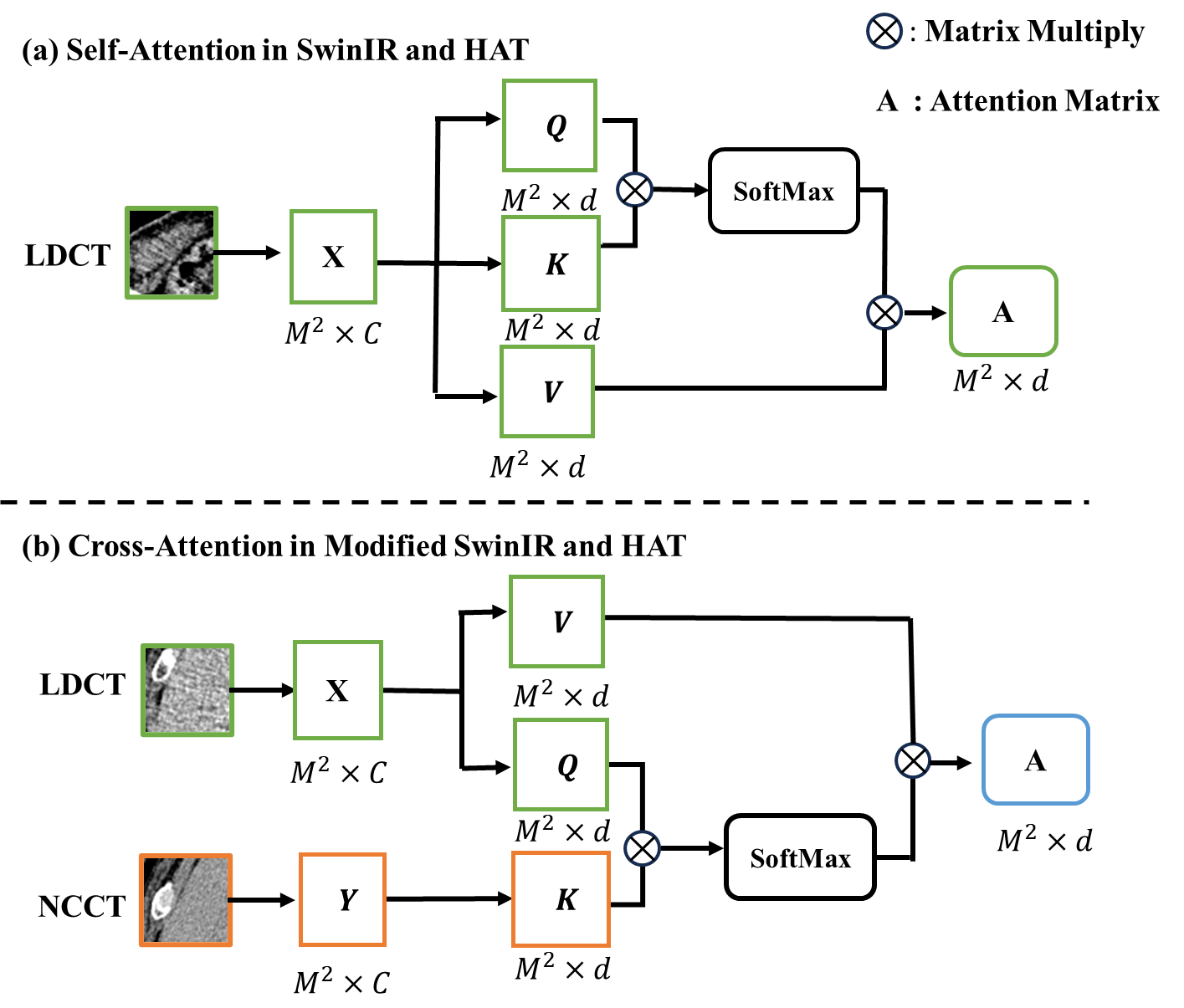}}
\caption{\textbf{Architectures of Self-Attention (SA) in vanilla SwinIR/HAT and Cross-Attention (CA) in modified SwinIR/HAT to incorporate the guidance of NCCT images}.
 %Based on this, we further propose the PTSP strategy to select image patches with high structural matching degree for training.
%The \textcolor[RGB]{12,140,218}{light blue} arrows indicate consistent structures or details of the images.
}
\label{fig:architecture}
\end{figure}

\noindent
\textbf{Similarity computation}. To select highly-similar triplets of image patches for network training, it is essential to measure the similarity between LDCT and NDCT patches as well as LDCT and NCCT patches in our PTSP strategy. The greater difference in pixel values between patches at a position should indicate lower similarity at that position. To this end, we utilize a set of weights $\{\pi_i\}_{i=0}^{n-1}$ to satisfy $0=\pi_{n-1}<\pi_{n-2}<...<\pi_{0}=1$. For example, when n=3, we can define $\{\pi_0=1, \pi_1=0.7, \pi_2=0\}$ to describe the differences in three levels. Denoting $\bm{S}_{LN}$ as the similarity mask to be computed, we assign different weights to different values of $\bm{D}_{LN}(x)$ and compute the similarity mask $\bm{S}_{LN}$ as follows:
\begin{equation}  
\bm{S}_{LN}(x) = 
\begin{cases}  
\pi_0, & \text{if  } \bm{D}_{LN}(x)=0,\\
\pi_{j}, & \text{if  } \bm{D}_{LN}(x)=j, 0<j<n-1,\\
\pi_{n-1}, & \text{if  } \bm{D}_{LN}(x)=n-1.
\end{cases}  
\label{discretization}  
\end{equation}
The similarity mask $\bm{S}_{LG}$ can be similarly defined. Then we compute the proportion of the sum of non-zero values in $\bm{S}_{LN}$ (or $\bm{S}_{LG}$) to the total number of pixels (\ie, $p^2$) in one LDCT image patch, as the mask similarity between LDCT and NDCT (or NCCT) image patches.

\noindent
\textbf{Selection of training patch triplets}. Here, the NCCT images are used to provide structural guidance for real-world LDCT image denoising. However, the structural misalignment between LDCT and NDCT images is inconsistent as that between LDCT and NCCT images. Therefore, when the similarity between LDCT and NDCT image patches reaches a preset threshold $s\in(0,1)$, we further search the surrounding area of the corresponding NCCT image patch to find the NCCT patch with the highest similarity to the LDCT patch. If the similarity between the LDCT and the NDCT image patch, as well as between the LDCT and the NCCT image patch, both exceed $s$, we will include this patch triplet into the training dataset. The threshold is set as $s=0.85$ in our experiments and can be adjusted flexibly based on the actual dataset.

With highly similar ``triplets'' of LDCT, NDCT, and NCCT image patches, we modify and train the denoising transformers with the supervision of NDCT images and the guidance of NCCT images. In our PTSP strategy, we set the threshold of mask similarity as $s=0.85$, which performs best in our ablation studies (\cref{subsec:ablation}).
To study the effectiveness of our PTSP strategy, we also construct a training dataset using ``pairs'' of LDCT and NDCT images. Here, we only need to compare the similarity between LDCT and NDCT image patches. The LDCT and corresponding NDCT image patches with a similarity over $s=0.85$ are included into the training set. We call this as Patch Similarity Purification (PSP) strategy~\cite{psp2024}.

\subsection{Training Denoising Transformers with NCCT Guidance}
\label{subsec:training}
Here, we modify two image denoising transformers of SwinIR~\cite{SwinIR} and HAT~\cite{chen2023hat} to exploit useful NCCT Guidance (NCG) for guided LDCT image denoising. For the networks without guidance, we only compute the similarity between the LDCT and corresponding NDCT image patches.

\noindent
\textbf{Modifying SwinIR~\cite{SwinIR} with our NCG for guided LDCT image denoising}.
%先说一下原有的swinir中的attention框架
Given an input feature of size $H \times W \times C$, the self-attention block in SwinIR first reshapes the feature into a size of $\frac{HW}{M^{2}} \times M^{2} \times C$ by partitioning it into non-overlapping $M \times M$ local windows, where $\frac{H W}{M^{2}}$ is the number of local windows. Then, self-attention is performed separately for the feature map in each local window $\bm{X} \in \mathbb{R}^{M^{2} \times C}$ (\figref{fig:architecture} (a)).
%the query, key and value matrices \xj{Q, K and V} are computed as: 
% \begin{align}
% Q= X P_{Q}, \quad K= X P_{K}, \quad V= X P_{V}
% \end{align}
%接着说我们的cross-attention
To incorporate the useful information of NCCT images for guided LDCT image denoising, we replace the self-attention in transformer layer of SwinIR by the cross-attention mechanism~\cite{Chen_2021_ICCV}.
As shown in \figref{fig:architecture} (b), we extract local window features $\bm{X}$ from LDCT image patch and $\bm{Y} \in \mathbb{R}^{M^{2} \times C}$ from the corresponding NCCT image patch. Here, the query, key, and value matrices $\bm{Q}$, $\bm{K}$, and $\bm{V}$ are computed as:
\begin{align}
\bm{Q}=\bm{X} \bm{P}_{Q}, \quad \bm{K}= \bm{Y} \bm{P}_{K}, \quad \bm{V}= \bm{X} \bm{P}_{V},
\end{align}
where $\bm{P}_{Q}$, $\bm{P}_{K}$, and $\bm{P}_{V}$ are linear projection matrices that are shared across different local windows. Then we have $\mathbf{Q}$, $\mathbf{K}$, and $\mathbf{V}$ all of size $M^2\times d$. The attention matrix is computed the same as that in self-attention:
\begin{align}
\text{Attention}(\bm{Q}, \bm{K}, \bm{V}) & = \operatorname{SoftMax}\left(\bm{Q} \bm{K}^{\top}/\sqrt{d}+\bm{B}\right) \bm{V},
\end{align}
where $\bm{B}$ is the learnable relative positional encoding.
For modified SwinIR, we use four RSTB layers with two Swin-Transformer layers in each RSTB layer.

\noindent
\textbf{Modifying HAT~\cite{chen2023hat} with our NCG for guided LDCT image denoising}. HAT combines channel attention~\cite{senet} and window-based self-attention for feature learning.
%What's more, to better aggregate the cross-window information, HAT introduces an overlapping cross-attention module to enhance the interaction between neighboring window features.
Here, we also replace the vanilla self-attention (\figref{fig:architecture} (a)) with cross-attention (\figref{fig:architecture} (b)) to utilize the NCCT images for useful guidance on LDCT image denoising. Similar to the modification on SwinIR, we extract local window features $\bm{X}$ and $\bm{Y}$ from the LDCT and NCCT image patches, respectively. We then transform $\bm{X}$ to the query/value matrices and transform $\bm{Y}$ to the key matrix. For modified HAT, we use four RHAG layers with two HAB blocks in each RHAG layer.

%which can lead to overly smooth highlights in the CT image and missing structural details in the edges. 
\noindent
\textbf{Loss function}. We train different denoising transformers with a combination of Charbonnier loss function $\ell_{C}$~\cite{charbonnier1994two} and perceptual loss function $\ell_{P}$~\cite{johnson2016perceptual}, as follows:
\begin{equation}
\label{eqn:loss}
\mathcal{L}=\ell_{C}+\lambda \ell_{P},
\end{equation}
where $\lambda$ is a hyper-parameter to trade-off the two loss functions. Here, we simply set $\lambda=1$.

\begin{figure}[t]
  \centering
  \begin{overpic}[width=0.24\textwidth]{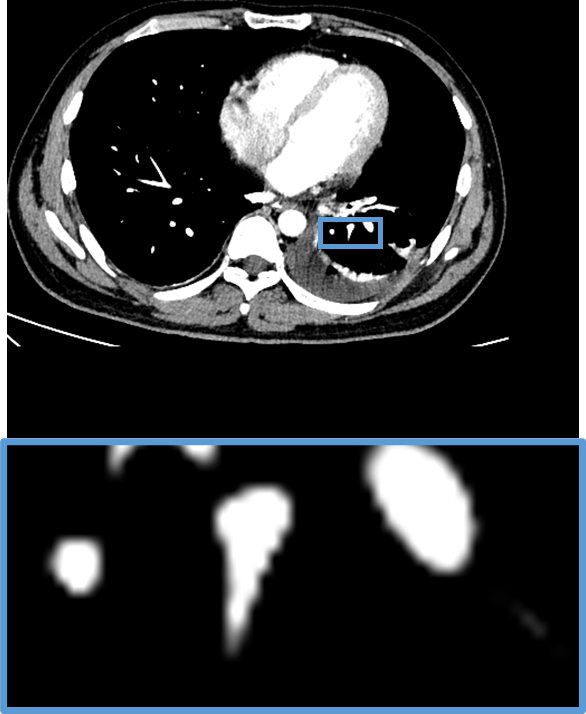}
    \put(-5,-8){\textbf{(a) Clinical NDCT Image}}
  \end{overpic}
  \begin{overpic}[width=0.24\textwidth]{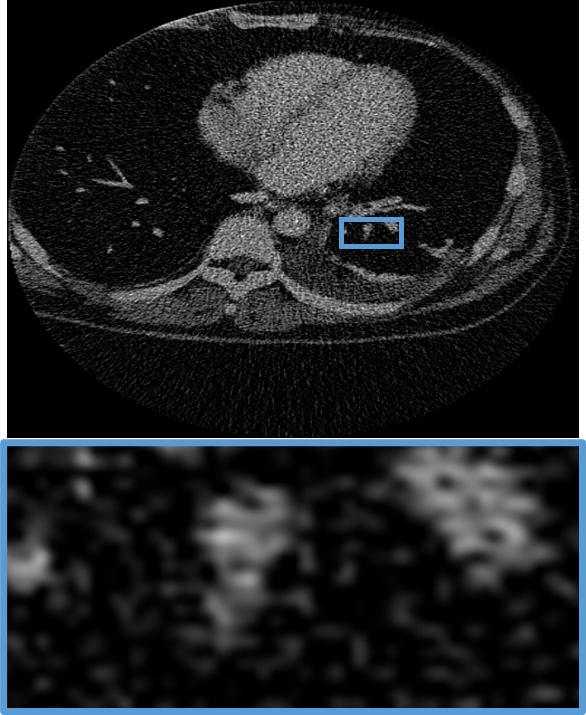}
    \put(-12,-8){\textbf{(b) Synthetic LDCT-Poisson noise}}
  \end{overpic}

  \vspace{15pt} % 调整垂直间距

  \begin{overpic}[width=0.24\textwidth]{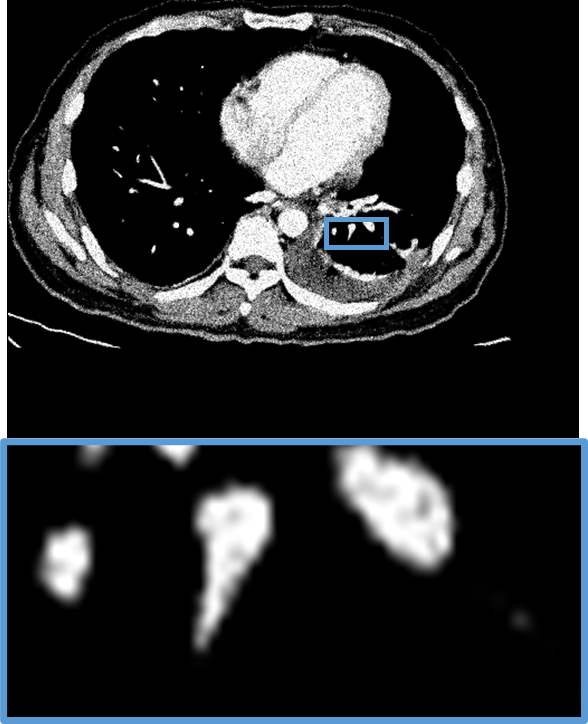}
    \put(5,-8){\textbf{(c) Our Synthetic LDCT}}
  \end{overpic}
  \begin{overpic}[width=0.24\textwidth]{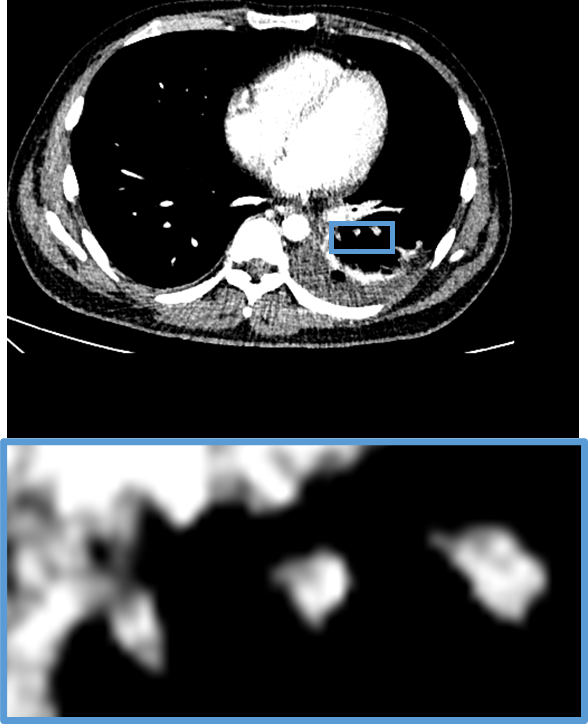}
    \put(2,-8){\textbf{(d) Clinical LDCT Image}}
  \end{overpic}
 \vspace{5pt}
  \caption{\textbf{Synthetic LDCT images based on the Poisson noise adding to the sinogram data of the corresponding NDCT image \textsl{v.s.} our synthetic LDCT image}. (a) NDCT image from real world. (b) Synthetic LDCT image by adding Poisson noise adding to the sinogram data of the corresponding NDCT image. (c) Our synthetic LDCT image. (d) Real-world LDCT image.}
  \label{fig:synthetic}
\end{figure}

\section{Our Datasets}
\label{sec:dataset}
\subsection{Synthetic Dataset}
Some studies~\cite{chen2017low} in the past synthesize the LDCT images by adding Poisson noise to the sinogram data of the NDCT ones. However, in the real world, there is not only noise between LDCT images and NDCT images, but also overall image shift and incomplete shape matching. Based on this, we introduce random displacement and elastic deformation~\cite{tanxing} when synthesizing LDCT images. The synthesis of LDCT images can be divided into the following three steps.

\textbf{Add random displacement}. Firstly, for the randomly divided test set, training set, and validation set of normal dose CT images, we randomly shift
the NDCT images of the divided training, validation, and test
sets by $2\sim 5$ pixels horizontally or vertically. 

\textbf{Add elastic deformation}. Secondly, we introduce elastic deformation~\cite{tanxing} to the image to simulate the structural distortion of low-dose CT images compared to normal-dose CT images in the real world. When adding elastic deformation, we set the control factor $\alpha$ as 25. 

\textbf{Add Gaussian noise}. Thirdly, we add zero-mean Gaussian noise with a standard deviation of $\sigma=40$ to generate synthetic LDCT images.
For each triplets of LDCT, NDCT, and NCCT patches, we crop it into $64\times 64$ patches. 

From \figref{fig:synthetic}, we can see that compared to add Poisson noise to the sinogram data of the NDCT images, our simulated LDCT images not only simulate the noise of clinical LDCT images, but also simulate the structural distortions between LDCT images and NDCT images in the real world.

\begin{figure*}[t]
  \begin{overpic}[width=0.33\textwidth]{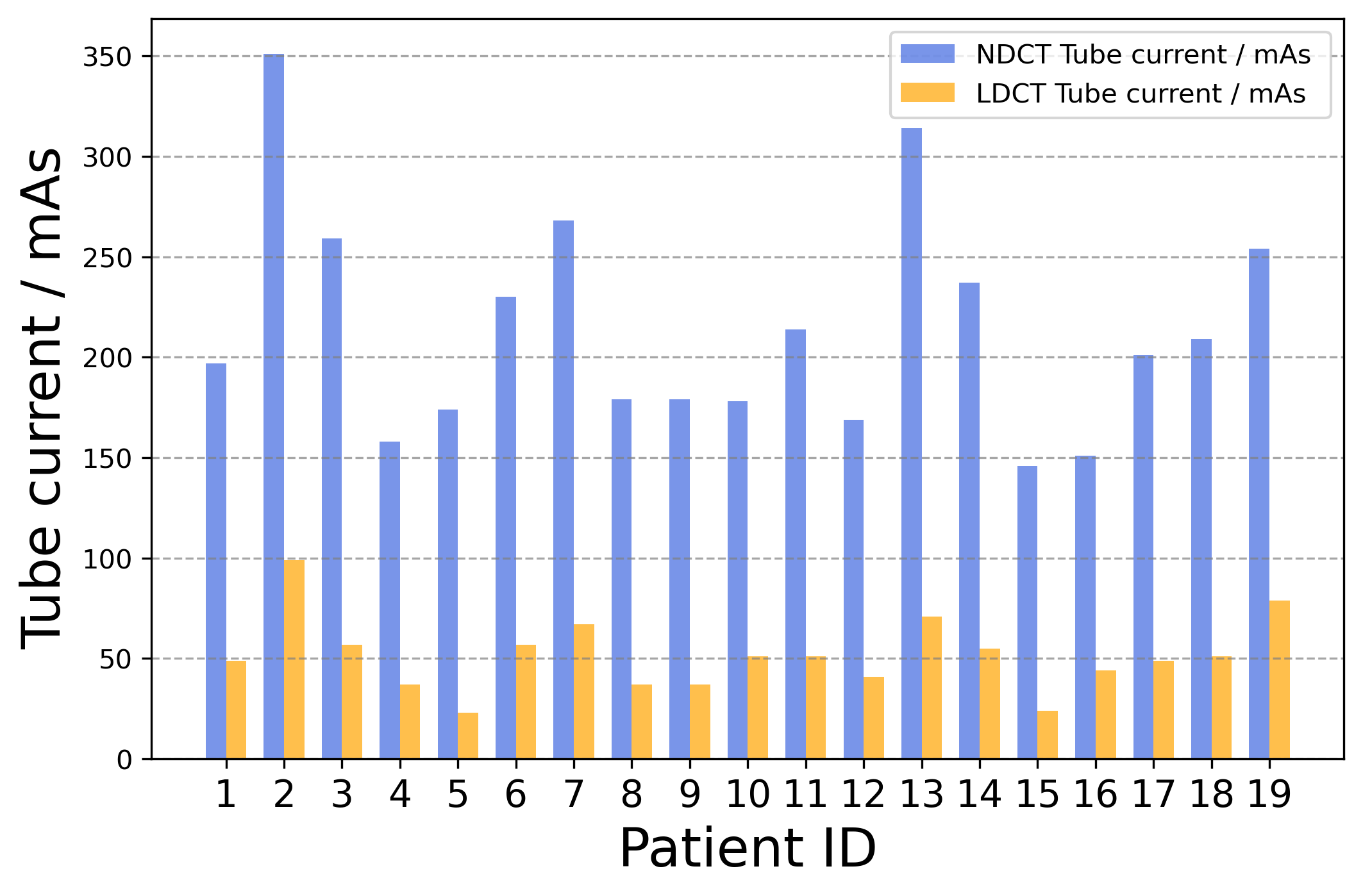}%
    \put(20,-8){(a) Scanning Tube Current}%
    % \label{fig1:LDCT}
  \end{overpic}%
  ~%
  \begin{overpic}[width=0.33\textwidth]{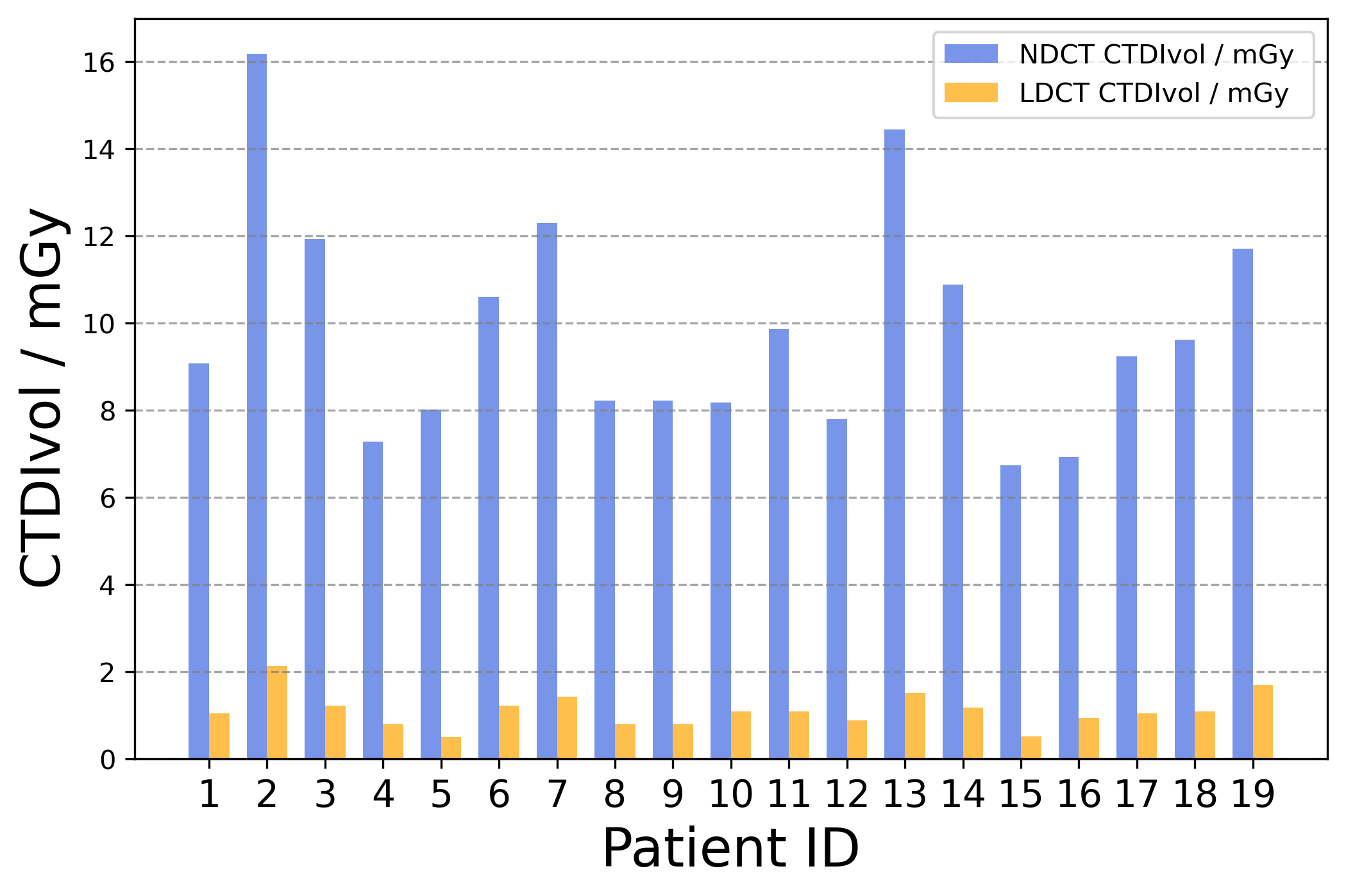}%
    \put(27,-8){(b) Patient's CTDIvol}%
    % \label{fig1:NDCT}
  \end{overpic}%
  ~%
  \begin{overpic}[width=0.33\textwidth]{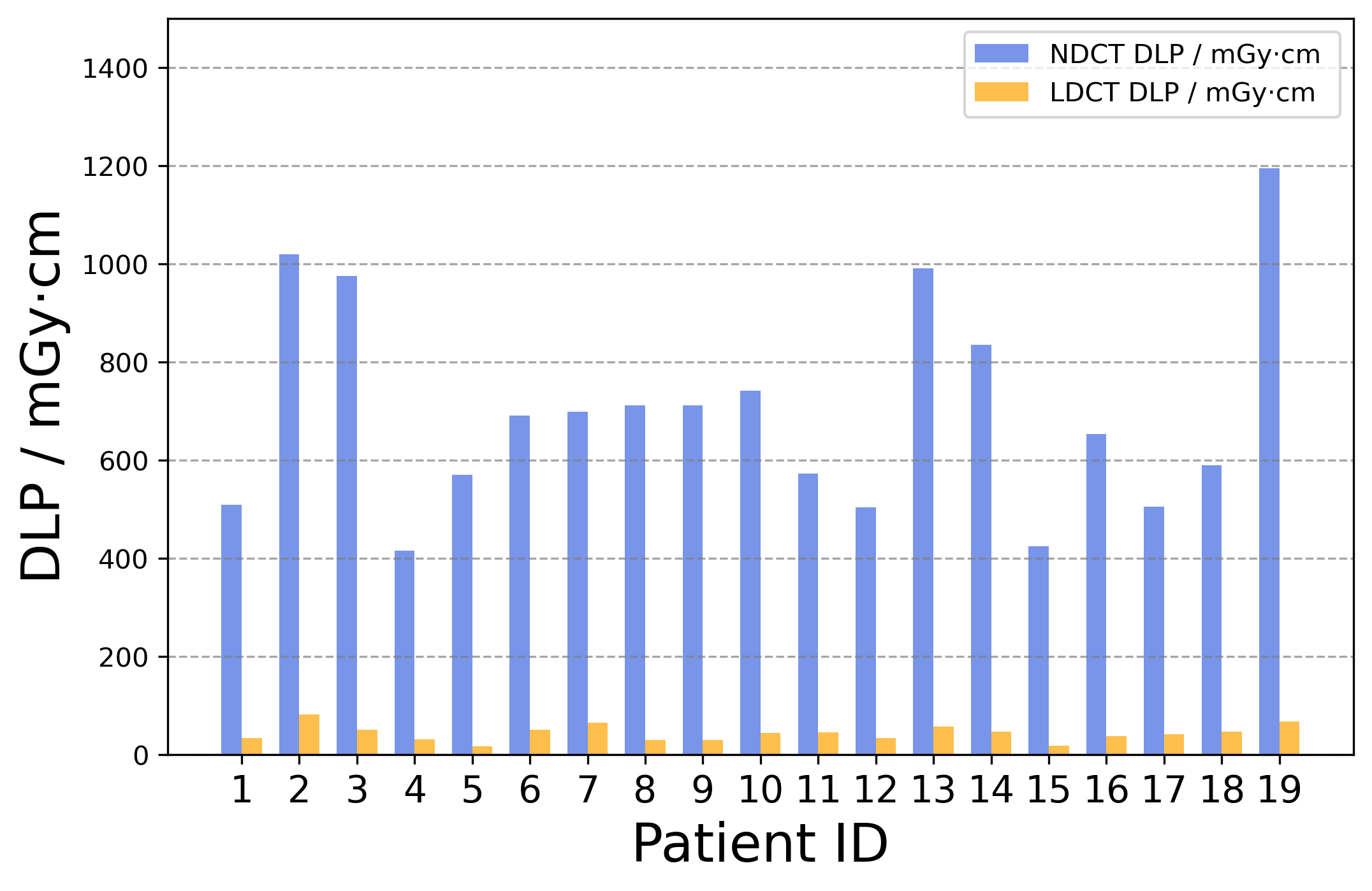}%
    \put(35,-8){(c) Patient's DLP}%
    % \label{fig1:mis}
  \end{overpic}%
   \vspace{10pt} % 调整此值来改变 caption 的垂直位置
\caption{\textbf{Statistics of the clinical dataset}. (a) The tube current parameters during the patient scanning process. The normal dose tube voltage is 100 kVp, while the low dose is 80 kVp. The NCCT is the same as the normal dose.
(b) Average CTDIvol values of the three scan phases for the patients.
(c) Average DLP values of the three scan phases for the patients.}
  \label{fig:CT}
  \vspace{-5mm}
\end{figure*}
\subsection{Clinical Dataset}
\subsubsection{Existing Datasets and Our Motivations}
Common CT noise reduction datasets include: the AAPM-Mayo dataset~\cite{aapm}, the NBIA/NCIA dataset~\cite{nbia}, the Piglet dataset~\cite{yi2018sharpness}, the Data Science Bowl 2017~\cite{yi2018sharpness}. The first two datasets synthesize the LDCT images from adding Poisson noise to the sinogram data of the NDCT ones. The AAPM-Mayo dataset comprises 2,378 512×512 NDCT images and simulated LDCT images
from 10 patients generated by adding Poisson noise in the sinogram domain of NDCT images~\cite{aapm}. In this dataset, the
radiation dose from the LDCT images are approximately 25\% of that from the NDCT images. The NBIA/NCIA dataset contains 7,015 NDCT images with diverse organ data~\cite{nbia}. However, the distribution of simulated noise is very different from the actual clinical noise. The Piglet dataset contains 850 pairs of LDCT and NDCT whole-body images, with four levels of noise by adjusting the tube current to 50\%, 25\%, 10\%, and 5\% of that used in capturing NDCT images~\cite{yi2018sharpness}. The Data Science Bowl
2017 provides 850 low-dose CT images from high-risk patients, with no corresponding NDCT ones~\cite{yi2018sharpness}. The two smaller datasets are appropriate for performance evaluation but unsuitable for training advanced networks. In summary, the existing CT noise reduction datasets face challenges of discrepancies between simulations and real-world scenarios, along with limited data volume. To address the challenges above, we have collected a large number of NCCT images and matched LDCT and NDCT images from three-phase scans from clinical practice.

\subsubsection{Introduction of Our clinical dataset}
The NCCT images refer to the internal cross-sectional images of the human body reconstructed by a computer based on the attenuation degree of X-rays, which are obtained by passing X-ray beams through the body from different angles without injecting contrast agents. They are usually used for preliminary examinations. The LDCT and NDCT images are obtained from three-phase scanning, which is a special CT examination method involving three different stages of CT scanning after intravenous injection of contrast agents.
The three phases are commonly referred to as the arterial phase, the venous phase or equilibrium phase, and the delayed phase. They further reveal the vascularization characteristics and enhanced lesion patterns, which helps more accurate diagnoses of disease.

Here we construct a new dataset scanned from 19 patients to provide large-scale real-world LDCT, NDCT and NCCT image triplets. The data collection for this study was approved by the Institutional Review Board of The First Affiliated Hospital of Xi’an Jiaotong University. All CT scans were performed in a 320-row spiral CT scanner uCT960+ from United Imaging Healthcare, with a rotation time of 0.5 s/rotation, a pitch of 0.9937, and a collimation width of 80 mm. The acquired raw data were transformed into the final CT images in the United Imaging Healthcare’s uInnovation-CT Explorer platform (R001). During the scanning process, the vascular monitoring scans employed contrast agent tracking technology, with the monitoring plane positioned at the descending aorta of rabbits and used a trigger threshold of 100 Hounsffeld Units (HU). Once the threshold was reached, the scan was triggered with a delay time of 12 seconds. For the arterial phase, the normal-dose and low-dose scans were conducted at approximately 12.0 and 15.4 seconds, respectively. For the portal venous phase, the normal-dose and low-dose scans were conducted
at approximately 28.0 and 31.4 seconds, respectively. For the delayed phase, the normal-dose and low-dose scans were conducted at approximately 40.0 and 43.4 seconds, respectively. Compared to NDCT images, NCCT images only lack the addition of contrast agent, while other aspects of the scanning protocol remain the same.

%描述一下图就行。
For each CT scan, we recorded the metrics of Volume CT Dose Index (CTDIvol) and Dose Length Product (DLP)~\cite{goldman2007principles} to measure the radiation dose. The Effective Dose(ED) is calculated by $ED=DLP\times k$, where k is the radiation dose conversion factor usually set as 0.015 mSv/(mGy·cm) for the abdomen~\cite{kalender2014dose}. As shown in \figref{fig:CT}, with the decrease in scanning tube current, the CTDI and DLP decrease from the normal doses of 9.85 mGy and 701.09 mGy·cm to 1.10 mGy and 43.71 mGy·cm, though bringing a significant amount of noise and artifacts.

%描述一下图就行。
Finally, we collect 17,541 LDCT images and corresponding NDCT/NCCT of $512\times512$ abdominal CT images. In subsequent experiments, we cropped the black area without content, resulting in an image size of $392\times512$.

\section{Experiments}
In this section, we first introduce our experimental setting in \cref{subsec:setting}. We further compare our methods with other denoising networks on our synthetic and real-world patient datasets in \cref{subsec:comparison method-S} and \cref{subsec:comparison method}, respectively. Finally, we study the hyper-parameters of the proposed methods in \cref{subsec:ablation}.

\label{sec:experiment}

\subsection{Experimental Setting}
\label{subsec:setting}

\subsubsection{Implementation Details}
\label{subsec:implementation details}
The modified SwinIR and HAT for LDCT image denoising are optimized by AdamW~\cite{loshchilov2018fixing} with $\beta_1=0.9$ and $\beta_2=0.99$. The learning rate is initialized as $2\times 10^{-4}$ and dynamically adjusted using the MultiStepLR strategy. The batch size is set as 32 in all experiments. We train all LDCT image denoising networks on an NVIDIA RTX 3090 GPU with 24GB memory. The window size is $M=8$.

\subsubsection{Evaluation Metrics}
Due to the misalignment between real-world LDCT and NDCT images, we employ feature-level metrics such as Fréchet Inception Distance (FID)~\cite{heusel2017gans}, Kernel Inception Distance (KID)~\cite{binkowski2018demystifying} and sFID~\cite{sfid} to objectively evaluate the distributional distance (in terms of diversity and visual quality) between denoised LDCT images and clean NDCT images from our clinical dataset.

\subsection{Comparison on Synthetic LDCT Images}
\label{subsec:comparison method-S}
\subsubsection{Synthetic Dataset}
Similar to clinical dataset, we select 13,704 image triplets from 15 randomly selected patients as the training set, 1,032 LDCT images from one patient as the validation set, and 2,805 LDCT images from the rest 3 patients as the test set. For each LDCT, NDCT, or NCCT image from our training set, we crop it into $64\times64$ patches with a 32-pixel overlap to train the LDCT image denoising networks. In this way, our synthetic training set has total 970,737 triplets of LDCT, NDCT, and NCCT patches.

\subsubsection{Comparison Methods}
On one hand, we conducted ablation experiments based on the introduction of the PTSP strategy and cross-attention HAT. From the~\tabref{tab:threshold_Sy} and the~\tabref{tab:HATSy}, the results indicate that the optimal denoising results are achieved when the similarity threshold is set to 80\% and the segmentation interval is set to [0,64,128,256], which validates the effectiveness of our PTSP strategy and NCG guidance via cross-attention. On the other hand, we compare the modified HAT with eleven other noise reduction models.

\subsubsection{Objective Results}
From the~\tabref{tab:Result_Sy}, we summarize the objective results on our synthetic dataset. One can see that trained with our PTSP strategy and the NCCT image guidance via cross-attention, the modified HAT outperforms its vanilla models and other comparison methods on LDCT image denoising. This validates the effectiveness of our PTSP strategy on filtering the training data and NCCT image guidance on the restoration~\cite{restoration} of image structure for LDCT image denoising.
\begin{table}[]  
\centering  
\caption{\textbf{Results of modified HAT~\cite{chen2023hat} using different similarity thresholds $s$ in our PTSP strategy} on our synthetic dataset. ``NCG'': NCCT image guidance.}
\label{tab:threshold_Sy} 
\begin{tabular}{cccccc}  
\toprule  
\textbf{Method} & \textbf{Thre.} $s$ & \textbf{sFID$\downarrow$} & \textbf{FID$\downarrow$} & \textbf{KID$\downarrow$}\\  
\midrule  
\multicolumn{1}{c}{\multirow{3}{*}{HAT+NCG+PTSP}} & 70\% &38.00  & 28.29 & 2.13 \\  
 &\cellcolor{gray!20}80\% &\cellcolor{gray!20}\textbf{37.47}  & \cellcolor{gray!20}\textbf{25.55} &\cellcolor{gray!20}\textbf{1.67} \\  
 & 90\% &58.59  & 55.21 & 5.27  \\   
\bottomrule  
\end{tabular}  
\end{table}

\begin{table}[t]    
\centering
\caption{\textbf{Results of HAT~\cite{chen2023hat} using different discretization intervals $n=2,3,4$} when implementing our PTSP strategy with NCCT image guidance in our synthetic dataset.}   
\label{tab:HATSy}    
\begin{tabular}{lcccc} % 指定四列，分别为左对齐、居中对齐、居中对齐、居中对齐    
\toprule % 顶部线    
\textbf{Threshold ($T$)} & \textbf{sFID$\downarrow$} & \textbf{FID$\downarrow$} & \textbf{KID$\times$100$\downarrow$} \\    
\midrule % 列标题下的线    
{[}0,32,256{]} &40.70 & 32.78 &2.57 \\
{[}0,64,256{]} &42.76  &35.65  &2.80  \\
{[}0,128,256{]} &39.73  &29.35 &  2.17  \\
{[}0,170,256{]} &38.08  &26.37  &1.84     \\
{[}0,192,256{]} & 38.02 &28.97  & 2.14    \\
%\rowcolor{gray!20}
{[}0,85,170,256{]} &38.50  & 30.35 &2.26  \\  
\rowcolor{gray!20}
{[}0,64,128,256{]} &\textbf{37.47}  & \textbf{25.55}& \textbf{1.67}  \\   
{[}0,32,64,128,256{]} &38.68  &28.31  &1.97    \\  
{[}0,64,128,192,256{]} &38.72 &30.99  &2.28    \\  
\bottomrule % 底部线    
\end{tabular}    
\end{table}

\begin{table}[t]    
\centering    
\caption{\textbf{Results of the comparison methods on our synthetic dataset}. When using the PTSP strategy, we set the similarity threshold as $s=0.85$. The number of segments is set as $n=3$. $\{T_i\}_{i=1}^n$ is set as $\{0,64,128,256\}$, and the weights are set as $\{1,0.7,0\}$.
``NCG'': NCCT image guidance.}
\label{tab:Result_Sy}    
\begin{tabular}{rrrrr} % 指定四列，分别为左对齐、居中对齐、居中对齐、居中对齐    
\toprule % 顶部线    
\textbf{Method} & \textbf{sFID$\downarrow$} & \textbf{FID$\downarrow$} & \textbf{KID$\times$100$\downarrow$}\\    
\midrule % 列标题下的线    
RED-CNN~\cite{chen2017low} &42.52 & 32.35&2.51   \\ 
DnCNN~\cite{zhang2017beyond} &38.35 &27.79 & 1.96  \\  
NAFNet~\cite{chen2022simple} &41.57 &36.60 &2.97   \\ 
CTformer~\cite{wang2023ctformer} &55.73 &68.10 &7.01    \\  
WGAN-VGG~\cite{yang2018low} &108.40 &106.05 &9.52  \\  
NAC~\cite{xu2020noisy}   &82.67   &99.43 &10.21        \\
BM3D~\cite{dabov2007image}   &117.64 &114.37 &  11.46      \\
MLEFGN~\cite{MLEFGN}   & 37.79 &26.18 &1.78       \\
SKWGIF~\cite{SKWGIF}   &108.52 & 122.46  &12.68        \\
\midrule
HAT~\cite{chen2023hat} &38.05 &27.08 & 1.84 \\ 
+ PSP~\cite{psp2024} &39.10 &28.74 &2.06    \\ 
\cellcolor{gray!20}+ NCG + PTSP&\cellcolor{gray!20}\textbf{37.47} &\cellcolor{gray!20}\textbf{25.55} &\cellcolor{gray!20}\textbf{1.67}  \\ 
%PTSP+NCG+HAT~\cite{chen2023hat}&(-24\%)\textbf{80.34} &(-17\%)34.44 & (-29\%)1.70 &(-19\%) 2.22 \\ 
\bottomrule % 底部线    
\end{tabular}    
\end{table}

\begin{table}[t]    
\centering    
\caption{\textbf{Results of the comparison methods on our clinical dataset}. When using the PTSP strategy, we set the similarity threshold as $s=0.85$. The number of segments is set as $n=3$. $\{T_i\}_{i=1}^n$ is set as $\{0,64,128,256\}$, and the weights are set as $\{1,0.7,0\}$.
``NCG'': NCCT image guidance.}
\label{tab:Result}    
\begin{tabular}{rrrr} % 指定四列，分别为左对齐、居中对齐、居中对齐、居中对齐    
\toprule % 顶部线    
\textbf{Method} & \textbf{sFID$\downarrow$} & \textbf{FID$\downarrow$} & \textbf{KID$\times$100$\downarrow$} \\    
\midrule % 列标题下的线    
RED-CNN~\cite{chen2017low} &84.79 &38.16 & 2.10 \\ 
DnCNN~\cite{zhang2017beyond} &88.78 &48.16 & 3.34  \\  
NAFNet~\cite{chen2022simple} &106.49 &74.88 & 6.14  \\ 
CTformer~\cite{wang2023ctformer} &209.20 &72.74 & 7.19  \\  
WGAN-VGG~\cite{yang2018low} &133.04 &72.22 & 3.81  \\  
NAC~\cite{xu2020noisy}   & 104.32  &61.84 &  4.46   \\
BM3D~\cite{dabov2007image}   &147.52 &96.69&   8.46    \\
MLEFGN~\cite{MLEFGN}   & 87.80 &46.59 &  3.41    \\
SKWGIF~\cite{SKWGIF}   & 123.25& 71.68  &   5.26        \\
AIIR-1   &110.55 & 96.86  &9.42         \\
AIIR-3   &93.48 & 56.99  &4.48       \\
AIIR-5   &86.52 & 46.27  &3.34        \\
KARL-9   &107.88 & 57.14  & 3.57         \\
\midrule
SwinIR~\cite{SwinIR} &89.20 & 50.36 & 3.62 \\  
+ RMSE &86.74 & 47.87 &3.41   \\
+ PSP~\cite{psp2024}&83.26 &36.32  &1.96   \\
\cellcolor{gray!20}+ NCG + PTSP &\cellcolor{gray!20}\textbf{81.11} &\cellcolor{gray!20}\textbf{32.29} & \cellcolor{gray!20}\textbf{1.42}  \\ 
\hdashline
% 加一条虚线
%PTSP+NCG+SwinIR~\cite{SwinIR} &(-9\%)81.11 &(-45\%)\textbf{32.29} & (-62\%)\textbf{1.42} & (-44\%)1.72 \\  
HAT~\cite{chen2023hat} &105.18 &41.35 & 2.40  \\ 
+ RMSE &81.59 &38.00  &2.12   \\
+ PSP~\cite{psp2024} &80.46 &35.99 &1.89    \\ 
\cellcolor{gray!20}+ NCG + PTSP&\cellcolor{gray!20}\textbf{80.34} &\cellcolor{gray!20}\textbf{34.44} &\cellcolor{gray!20}\textbf{1.70}  \\ 
%PTSP+NCG+HAT~\cite{chen2023hat}&(-24\%)\textbf{80.34} &(-17\%)34.44 & (-29\%)1.70 &(-19\%) 2.22 \\ 
\bottomrule % 底部线    
\end{tabular}    
\end{table}

\subsection{Comparison on Real-World LDCT Images} 
\label{subsec:comparison method}
\subsubsection{Comparison Methods}
To study the effectiveness of our PTSP strategy and NCCT image guidance, we compare SwinIR~\cite{SwinIR} and HAT~\cite{chen2023hat} trained with or without our PTSP strategy on LDCT image denoising. Note that these methods using our PTSP need the incorporation of cross-attention to accommodate the NCCT image guidance.
We also compare the modified SwinIR and HAT trained with our PTSP strategy with fifteen image denoising methods, which can be divided into five categories: 1) three LDCT image denoising methods of WGAN-VGG~\cite{yang2018low}, RED-CNN~\cite{chen2017low}, and CTformer~\cite{wang2023ctformer}; 2) four image denoising methods of BM3D~\cite{dabov2007image}, DnCNN~\cite{zhang2017beyond}, Noise-As-Clean (NAC)~\cite{xu2020noisy}, and NAF-Net~\cite{chen2022simple}; 3) two baselines of SwinIR~\cite{SwinIR} and HAT~\cite{chen2023hat}; 4) two guided image denoising methods of MLEFGN~\cite{MLEFGN} and SKWGIF~\cite{SKWGIF}; 5) four commercial algorithms of AIIR-1, AIIR-3, AIIR-5, and KARL9 provided by United Imaging Healthcare.

\begin{figure*}[t]
  \centering

  \begin{minipage}[b]{0.194\textwidth}
    \begin{overpic}[width=\textwidth]{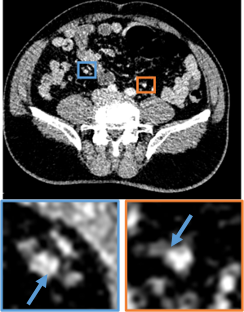}
      \put(-1,-10){ (a) \textbf{The LDCT Image}}
    \end{overpic}
  \end{minipage}
    \begin{minipage}[b]{0.194\textwidth}
    \begin{overpic}[width=\textwidth]{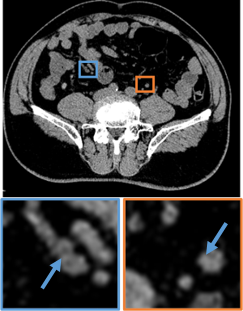}
      \put(1,-10){(b) \textbf{The NCCT Image}}
    \end{overpic}
  \end{minipage}
    \begin{minipage}[b]{0.194\textwidth}
    \begin{overpic}[width=\textwidth]{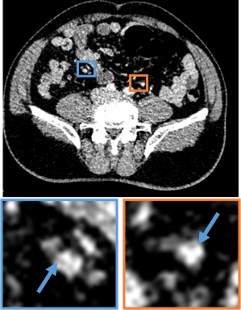}
      \put(15,-10){(c) \textbf{NAC~\cite{xu2020noisy}}}
    \end{overpic}
  \end{minipage}
    \begin{minipage}[b]{0.194\textwidth}
    \begin{overpic}[width=\textwidth]{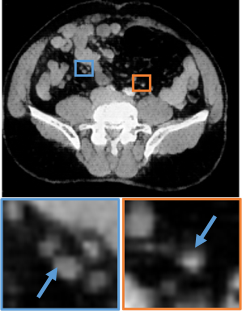}
      \put(12,-10){(d) \textbf{NAFNet~\cite{chen2022simple}}}
    \end{overpic}
  \end{minipage}
    \begin{minipage}[b]{0.194\textwidth}
    \begin{overpic}[width=\textwidth]{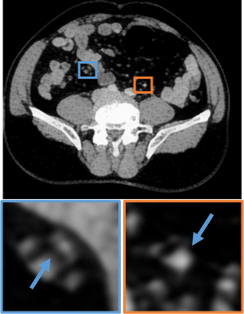}
      \put(13,-10){(e) \textbf{DnCnn~\cite{zhang2017beyond}}}
    \end{overpic}
  \end{minipage}
  \vspace{2.0em}
  
    \begin{minipage}[b]{0.194\textwidth}
    \begin{overpic}[width=\textwidth]{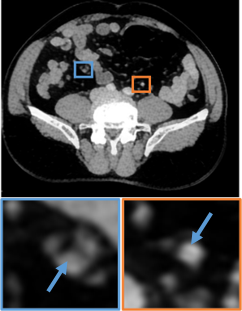}
      \put(9,-10){(f) \textbf{MLEFGN~\cite{MLEFGN}}}
    \end{overpic}
  \end{minipage}
    \begin{minipage}[b]{0.194\textwidth}
    \begin{overpic}[width=\textwidth]{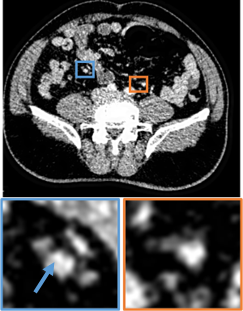}
      \put(8,-10){(g) \textbf{SKWGIF~\cite{SKWGIF}}}
    \end{overpic}
  \end{minipage}
    \begin{minipage}[b]{0.194\textwidth}
    \begin{overpic}[width=\textwidth]{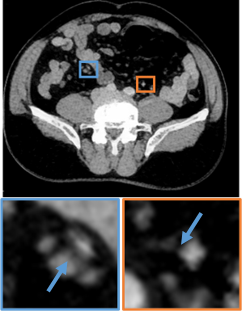}
      \put(8,-10){(h) \textbf{RED-CNN~\cite{chen2017low}}}
    \end{overpic}
  \end{minipage}
    \begin{minipage}[b]{0.194\textwidth}
    \begin{overpic}[width=\textwidth]{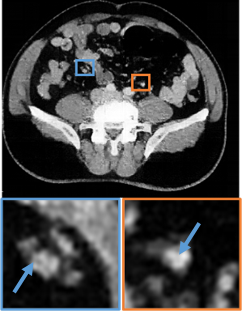}
      \put(8,-10){(i) \textbf{CTformer~\cite{wang2023ctformer}}}
    \end{overpic}
  \end{minipage}
      \begin{minipage}[b]{0.194\textwidth}
    \begin{overpic}[width=\textwidth]{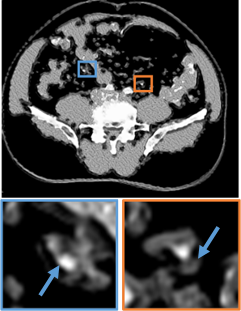}
      \put(3,-10){(j) \textbf{WGAN-VGG~\cite{yang2018low}}}
    \end{overpic}
  \end{minipage}
  \vspace{2.0em}
  
    \begin{minipage}[b]{0.194\textwidth}
    \begin{overpic}[width=\textwidth]{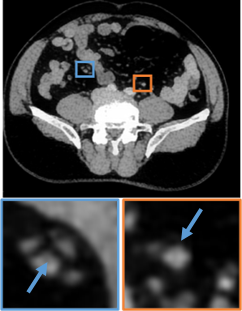}
      \put(13,-10){(k) \textbf{SwinIR~\cite{SwinIR}}}
    \end{overpic}
  \end{minipage}
  \begin{minipage}[b]{0.194\textwidth}
    \begin{overpic}[width=\textwidth]{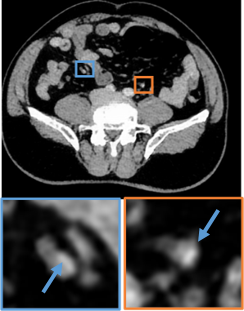}
      \put(-2,-10){(l) \textbf{SwinIR+NCG+PTSP}}
    \end{overpic}
  \end{minipage}
  \begin{minipage}[b]{0.194\textwidth}
    \begin{overpic}[width=\textwidth]{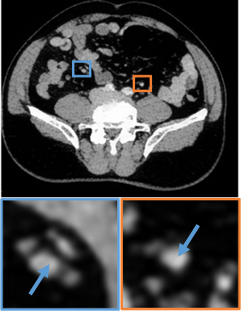}
      \put(15,-10){(m) \textbf{HAT~\cite{chen2023hat}}}
    \end{overpic}
  \end{minipage}  
    \begin{minipage}[b]{0.194\textwidth}
    \begin{overpic}[width=\textwidth]{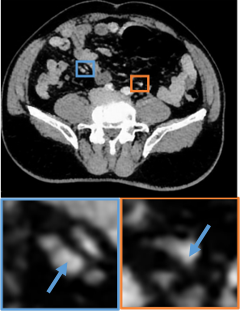}
      \put(1,-10){(n) \textbf{HAT+NCG+PTSP}}
    \end{overpic}
  \end{minipage}
    \begin{minipage}[b]{0.194\textwidth}
    \begin{overpic}[width=\textwidth]{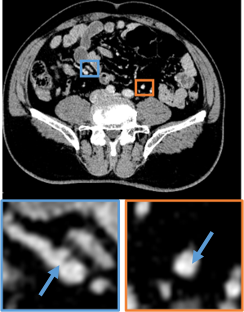}
      \put(2,-10){(o) \textbf{The NDCT Image}}
    \end{overpic}
%  \end{minipage}
%    \begin{minipage}[b]{0.194\textwidth}
%    \begin{overpic}[width=\textwidth]{ljh/SwinIR.png}
%      \put(2,-10){(i) \textbf{SwinIR}}
%    \end{overpic}
%  \end{minipage}
%      \begin{minipage}[b]{0.194\textwidth}
%    \begin{overpic}[width=\textwidth]{ljh/guided swinir.png}
%      \put(3,-10){(j) \textbf{Guided SwinIR}}
%    \end{overpic}
  \end{minipage}
  \vspace{1.0em}
    \caption{\textbf{Comparison of visual quality by different denoising methods} on one LDCT image from our clinical dataset.}
  \label{fig:methods}
  \vspace{-3mm}
\end{figure*}

\subsubsection{Objective Results}
Here, we summarize the results of comparison methods on the test set of our clinical dataset. The results summarized in the~\tabref{tab:Result} show that, trained with our PTSP strategy and the cross-attention for the guidance of NCCT images, the modified SwinIR and HAT outperforms not only their vanilla models but also the other comparison methods on LDCT image denoising, in terms of the sFID, FID, and KID.
%
%Besides, \xj{SwinIR and HAT with PSP?}
Besides, trained with PSP strategy~\cite{psp2024}, SwinIR and HAT outperform the vanilla models, in terms of the sFID, FID, and KID. 
%最后来一个总结
This indicates that introducing NCCT image guidance and our PTSP strategy effectively improves the denoising effect of these methods.

\subsubsection{Visual Quality}
We compare the visual results of different image denoising methods on our clinical dataset. As shown in~\figref{fig:methods}, the denoised image obtained by BM3D is over-smooth. The results of NAC and SKWGIF exhibit weak denoising effects. WGAN-VGG not only removes the noise but also generates visual artifacts, which would have significant side impacts on clinical diagnosis. The results of NAFNet, CTFormer, and DnCNN have obvious motion artifacts with some blurry areas. The results of RED-CNN and MLEFGN are close to the NDCT images from the overall visual effect, but suffer from distorted or blurry structure.
Additionally, both the vanilla SwinIR and HAT exhibit slight motion artifacts and inconsistent details with the NDCT images.
With our PTSP strategy, the modified SwinIR and HAT not only obtain results close to the NDCT images from the overall visual effect, but also well preserve the structure and details of the LDCT image.

\subsection{Ablation Study}
\label{subsec:ablation}
We conduct ablation studies to explore the working mechanism of our PTSP strategy and NCCT-guided cross-attention. Specifically, we assess: 
1) the influence of different attention mechanisms to SwinIR and HAT on LDCT image denoising; 
2) how do the number of segments $n$ and the segmentation points $\{T_i\}_{i=0}^n$ affect the size of training set and the performance of modified SwinIR and HAT on guided LDCT image denoising; 
3) the impact of different similarity thresholds $s$ for guided LDCT image denoising.

\noindent
\textbf{1) The influence of different attention mechanisms to SwinIR and HAT on LDCT image denoising}.
The vanilla SwinIR and HAT use the self-attention (SA) mechanism. To incorporate the guidance from NCCT images, we modify SwinIR and HAT with our NCCT image guidance (NCG), which is implemented by replacing the SA with cross-attention (CA). To study its effectiveness, we compare the denoising results of SwinIR and HAT with or without our NCG on the training set selected by our PTSP strategy.
From the~\tabref{tab:attention}, one can see that after introducing NCCT image guidance (NCG), the modified SwinIR and HAT achieve boosted results on all metrics. For example, modified SwinIR achieves an improvement of sFID, FID, and KID by 2.6\%, 11.1\%, and 27.6\%, respectively, while modified HAT achieves consistent improvements on all metrics by 0.11\%, 5.4\%, and 12.2\%, respectively. This demonstrates that after introducing NCCT image guidance, the modified SwinIR and HAT not only effectively remove the realistic CT noise but also well preserve the LDCT image structure.

\begin{table}[t]  
\centering  
\caption{\textbf{Results of SwinIR~\cite{SwinIR} and HAT~\cite{chen2023hat} using self-attention (SA) or cross-attention (CA) mechanism} on our clinical dataset. For SA mechanism, the data screening strategy is the PSP strategy~\cite{psp2024} since there is no NCCT image guidance.}
\label{tab:attention} 
\begin{tabular}{lccccc}  
\toprule  
\textbf{Method}&\textbf{Attention}  &\textbf{sFID$\downarrow$} & \textbf{FID$\downarrow$} & \textbf{KID$\downarrow$} \\  
\midrule  
SwinIR+PSP~\cite{psp2024} & SA & 83.27 & 36.32 & 1.96 \\  
SwinIR+NCG+PTSP & \cellcolor{gray!20}CA & \cellcolor{gray!20}\textbf{81.11} & \cellcolor{gray!20}\textbf{32.29} & \cellcolor{gray!20}\textbf{1.42} &  \\  
% & self+cross&PTSP & \textbf{80.92} & 35.40 & 1.94 & 1.96 \\  
\midrule  
HAT+PSP~\cite{psp2024} & SA & 80.45 & 35.99 & 1.89\\  
 HAT+NCG+PTSP& \cellcolor{gray!20}CA&\cellcolor{gray!20}\textbf{80.34} & \cellcolor{gray!20}\textbf{34.05} & \cellcolor{gray!20}\textbf{1.66}  \\  
 % & self+cross&PTSP & 81.14 & 35.50 & 1.85 & 2.28 \\  
\bottomrule  
\end{tabular}  
\end{table}

\begin{table}[t]    
\centering    
\caption{\textbf{Results of SwinIR~\cite{SwinIR} using different discretization intervals $n=2,3,4$}  when implementing our PTSP strategy with NCCT image guidance in our clinical dataset.}    
\label{tab:SwinIR}    
\begin{tabular}{lcccc} % 指定四列，分别为左对齐、居中对齐、居中对齐、居中对齐    
\toprule % 顶部线    
\textbf{Threshold ($T$)} & \textbf{sFID$\downarrow$} & \textbf{FID$\downarrow$} & \textbf{KID$\times$100$\downarrow$} \\    
\midrule % 列标题下的线     
{[}0,32,256{]} & 84.66 & 36.04 & 1.86 \\
{[}0,64,256{]} &84.10  & 35.63 & 1.71\\
{[}0,128,256{]} &88.47  &42.54 & 2.49  \\
{[}0,170,256{]} &90.69  & 37.47 & 1.54   \\
{[}0,192,256{]} & 86.06 &37.30  &1.74     \\
%\rowcolor{gray!20}
{[}0,85,170,256{]} &81.50 & \textbf{32.26} & \textbf{1.36} \\  
\rowcolor{gray!20}
{[}0,64,128,256{]} &\textbf{81.11} & 32.29 & 1.42  \\   
{[}0,32,64,128,256{]} &83.04  &33.65  &1.56   \\  
{[}0,64,128,192,256{]} &81.87  & 33.04 & 1.39 \\  
\bottomrule % 底部线    
\end{tabular}    
\end{table}

\begin{table}[t]    
\centering
\caption{\textbf{Results of HAT~\cite{chen2023hat} using different discretization intervals $n=2,3,4$}  when implementing our PTSP strategy with NCCT image guidance in our clinical dataset.}   
\label{tab:HAT}    
\begin{tabular}{lcccc} % 指定四列，分别为左对齐、居中对齐、居中对齐、居中对齐    
\toprule % 顶部线    
\textbf{Threshold ($T$)} & \textbf{sFID$\downarrow$} & \textbf{FID$\downarrow$} & \textbf{KID$\times$100$\downarrow$} \\    
\midrule % 列标题下的线    
{[}0,32,256{]} &82.14 & 35.86 &1.70   \\
{[}0,64,256{]} &83.17  &34.48  &1.59 \\
{[}0,128,256{]} &88.41  &39.57 &  2.14  \\
{[}0,170,256{]} &85.68  &38.10  &1.59     \\
{[}0,192,256{]} & 85.04 &37.64  & 1.87    \\
%\rowcolor{gray!20}
{[}0,85,170,256{]} &81.73  & \textbf{33.83} & \textbf{1.47}  \\  
\rowcolor{gray!20}
{[}0,64,128,256{]} &\textbf{80.34}  & 34.05& 1.66 \\   
{[}0,32,64,128,256{]} &82.76  &35.79  &1.66   \\  
{[}0,64,128,192,256{]} &85.41 &41.57  &2.28   \\  
\bottomrule % 底部线    
\end{tabular}    
\end{table}

\begin{table}[t]  
\centering  
\caption{\textbf{Results of modified SwinIR~\cite{SwinIR} and HAT~\cite{chen2023hat} using different similarity thresholds $s$ in our PTSP strategy} on our clinical dataset. ``NCG'': NCCT image guidance.}
\label{tab:threshold} 
\begin{tabular}{cccccc}  
\toprule  
\textbf{Method} & \textbf{Thre.} $s$ & \textbf{sFID$\downarrow$} & \textbf{FID$\downarrow$} & \textbf{KID$\downarrow$} \\  
\midrule  
\multicolumn{1}{c}{\multirow{3}{*}{SwinIR+NCG+PTSP}} & 80\% &\textbf{80.43}  & 34.11 & 1.73  \\  
 &\cellcolor{gray!20}85\% &\cellcolor{gray!20}81.11  & \cellcolor{gray!20}\textbf{32.29} &\cellcolor{gray!20}\textbf{1.42}  \\  
 & 90\% &85.19  & 34.21 & 1.55  \\  
\midrule  
\multicolumn{1}{c}{\multirow{3}{*}{HAT+NCG+PTSP}} & 80\% &80.53 & 39.33 & 2.25 \\  
 & \cellcolor{gray!20}85\% &\cellcolor{gray!20}\textbf{80.34}  &\cellcolor{gray!20}\textbf{34.05} &\cellcolor{gray!20}\textbf{1.66} \\  
 & 90\% & 84.31 & 36.63 & 1.95\\  
\bottomrule  
\end{tabular}  
\end{table}

\noindent
\textbf{2) How do the number of segments $n$ and the segmentation points $\{T_i\}_{i=0}^n$ affect the size of training set and the performance of modified SwinIR and HAT on guided LDCT image denoising}?
As mentioned in \cref{subsec:PTSP}, our PTSP strategy requires to pre-define a discrete interval number $n$ and a set of discrete interval segmentation points $\{T_i\}_{i=0}^n$. The purpose of our PTSP strategy is to select highly similar patch triplets from the aligned LDCT images, NDCT images, and NCCT images. The role of $n$ and $\{T_i\}_{i=0}^n$ in this process is crucial, as they directly affect the quality of the training data as well as the denoising effect. Therefore, we set $n=2,3,4$ with different sets of $\{T_i\}_{i=0}^n$ to explore the appropriate number of segmentation points and suitable segmentation intervals on the modified SwinIR and HAT using NCG guidance and our PTSP strategy. From the~\tabref{tab:SwinIR} and the~\tabref{tab:HAT}, we observe that modified SwinIR and HAT achieve the best FID and KID results when $n=3$ and $\{T_i\}_{i=0}^3=[0,85,170,256]$ and achieve the best sFID results when $n=3$ and $\{T_i\}_{i=0}^3=[0,64,128,256]$. However, the performance degrades when $n$ is further increased to 4. Thus, we set $n=3$ and $\{T_i\}_{i=0}^3=[0,64,128,256]$ in our PTSP strategy.

\noindent
\textbf{3) The impact of different similarity thresholds $s$ for guided LDCT image denoising}. The hyperparameter of similarity threshold $s$ in our PTSP strategy mainly influences the size and quality of the training data. Higher threshold indicates better similarity quality but also decreases the number of patch triplets for network training. To choose a proper threshold for the modified SwinIR and HAT, we perform experiments on LDCT image denoising by setting $s=0.80$, $0.85$, and $0.90$ while keeping all other settings the same.
From the~\tabref{tab:threshold}, it can be seen that when the similarity threshold $s$ is set as 0.85, the modified SwinIR (SwinIR+NCG) achieves the best results on FID and KID, as well as the second best results on sFID (only 0.68 worse than that using the threshold of $s=0.80$). The modified HAT (HAT+NCG) obtains the best results on sFID, FID, and KID when the similarity threshold is set as $s=0.85$. These results demonstrate that smaller similarity threshold will bring more training data with stronger structure misalignment, resulting in worse denoising performance. A larger similarity threshold will lead to less training data, resulting in inadequate network training with degraded denoising performance. Therefore, it is proper to set the similarity threshold as $s=0.85$ in our PTSP strategy.

\section{Conclusion}
\label{sec:conclusion}

In this paper, by observing that the non-contrast CT (NCCT) images share similar context characteristics to the corresponding NDCT images from three-phase scanning, we proposed to incorporate useful information from clean NCCT images as useful guidance for real-world LDCT images denoising. We modified two image denoising transformers, \ie, SwinIR and HAT, by replacing the vanilla self-attention mechanism with the cross-attention mechanism to accommodate the NCCT image guidance. To alleviate the issue of spatial misalignment between real-world LDCT images and NDCT (or NCCT) images, we proposed a Patch Triplet Similarity Purification (PTSP) strategy to select highly similar triplets of LDCT, NDCT, and NCCT image patches for network training. Through extensive experiments on our clinical dataset, the modified SwinIR and HAT outperform fifteen comparison methods on LDCT image denoising. Extensive experiments on our clinical dataset demonstrate that the modified SwinIR and HAT outperform fifteen comparison methods on LDCT image denoising. They not only effectively remove the noise from LDCT images, but also preserve the original structure of LDCT images with the help of NCCT image guidance.

In the future, we will study the effectiveness of our NCCT image guidance and PTSP strategy on LDCT images with much lower radiation doses to pursue further radiation reduction. We hope that our work will be helpful for clinical applications.

\section{Acknowledgments}
The authors thank Liqi Xue for her invaluable contributions in helpful discussions and beautification of the figures.
       {\small 
		\bibliographystyle{ieee_fullname}

  }

\end{document}